\documentclass[preprint,review,12pt,3p]{elsarticle}
\usepackage{lineno,hyperref}
\modulolinenumbers[5]
\pdfoutput=1
\journal{}

\usepackage{amsmath}									
\usepackage{float}
\usepackage{graphicx}
\usepackage{subfigure}
\usepackage{dsfont}
\usepackage{amssymb}
\usepackage{natbib}
\begin{document}

\begin{frontmatter}

\title{Energy of low angle grain boundaries based on continuum dislocation structure}

\author[mymainaddress]{Luchan Zhang}
\author[mysecondaryaddress]{Yejun Gu}
\author[mymainaddress]{Yang Xiang\corref{mycorrespondingauthor}}
\cortext[mycorrespondingauthor]{Corresponding author}
\ead{maxiang@ust.hk}

\address[mymainaddress]{Department of Mathematics, The Hong Kong University of Science and Technology, Clear Water Bay, Kowloon, Hong Kong}
\address[mysecondaryaddress]{Nano Science and Technology Program, The Hong Kong University of Science and Technology, Clear Water Bay, Kowloon, Hong Kong}

\begin{abstract}

In this paper,  we present a continuum model to compute the energy of low angle grain boundaries for any given degrees of freedom (arbitrary rotation axis, rotation angle and boundary plane orientation) based on a continuum dislocation structure. In our continuum model, we minimize the grain boundary energy associated with the dislocation structure subject to the constraint of Frank's formula for dislocations with all possible Burgers vectors. This constrained minimization problem is solved by the penalty method by which it is turned into an unconstrained minimization problem. The grain boundary dislocation structure is approximated by a network of straight dislocations that predicts the energy and dislocation densities of the grain boundaries. The grain boundary energy based on the calculated dislocation structure is able to incorporate its anisotropic nature. We use our continuum model to systematically study the energy of $<111>$ low angle grain boundaries in fcc Al with  any boundary plane orientation and all six possible Burgers vectors. Comparisons with atomistic simulations results show that our continuum model is able to give excellent predictions of the energy and dislocation densities of low angle grain boundaries. We also study the energy of low angle grain boundaries in fcc Al with varying rotation axis while the remaining degrees of freedom are fixed.
With modifications, our model can also apply to dislocation structures and energy of heterogeneous interfaces.

\end{abstract}


\end{frontmatter}


\section{Introduction}

Energetic and dynamic properties of grain boundaries play vital roles in the mechanical and plastic behaviors of polycrystalline materials \cite{Sutton1995}. These properties of grain boundaries such as grain boundary energy and mobility strongly depend on the structures of grain boundaries, and have attracted  considerable research attention for many decades \cite{Sutton1995,ReadShockley1950,BRANDON1964813,hasson1971,GLEITER198291,SuttonVitek1983,SchwartzVitek1985,vitek1987,
WOLF1990781,WOLF1990791,Rittner1996,Saylor20033663,Saylor20033675,Olmsted20093694,Lim20095013,Holm2010905,Rohrer2011,
Holm20115250,Lim20121395,Wu2012407,Dai20131327,Winther2013,Bulatov2014161,Dai2014162,Shen2014125,Zhu2014175,Winther2014}.

Grain boundary energy and structure are determined by five macroscopic degrees of freedom (DOFs) that include the grain misorientation (three DOFs) and the boundary plane orientation (two DOFs) \cite{Sutton1995}. Early works focused on the grain boundary energy as a function of misorientation angle. In the classical theory of Read and Shockley \cite{ReadShockley1950},
the grain boundary energy is $E=E_0\theta(A-\ln\theta)$, where $\theta$ is the misorientation angle and parameters $E_0$ and $A$ depend on the grain boundary orientation. This energy formula was derived based on a dislocation model of grain boundaries with  cancellation of the long-range elastic fields.
 Hasson and Goux \cite{hasson1971} measured the energy of tilt boundaries in aluminum using both experimental method and atomistic calculation.
Wolf \cite{WOLF1990781,WOLF1990791} invested the structure-energy correlation of grain boundaries of different boundary planes by using molecular dynamics simulations, and described the energy as a function of the misorientation angle. A linear correlation between the energy and the grain boundary volume expansion was observed.
There were also studies of  grain boundary structure and energy based on coincidence site lattice (CSL) and displacement shift complete (DSC) dislocations or polyhedral/structure unit models  \cite{BRANDON1964813,GLEITER198291,SuttonVitek1983,SchwartzVitek1985}.

More systematic examinations of the dependence of the grain boundary energy on all five DOFs have been conducted in recent years.
 Olmsted \textit{et al.} \cite{Olmsted20093694,Holm2010905} calculated the energies of a set of 388 distinct grain boundaries by using atomistic simulations, and examined the correlations of the boundary energy with other boundary properties. Especially, they showed that the grain boundary plane orientation is crucial in the determination of boundary energy. Saylor \textit{et al.} and Holm \textit{et al.} \cite{Saylor20033663,Saylor20033675,Holm20115250} experimentally measured the grain boundary character distributions, and then used the character distributions to reconstruct the grain boundary energies with fixed rotation axes and misorientation angles.
They also compared these experimental results of energies with those computed by using molecular dynamics simulations \cite{Holm20115250}. Assuming the grain boundary energy is a continuous function of the five DOFs, Bulatov \textit{et al.} \cite{Bulatov2014161} constructed a closed-form grain boundary energy function for fcc metals by using a hierarchical interpolation from values of special sets of grain boundaries (low dimensional subsets termed "grofs").

Low angle grain boundaries can be modelled as arrays of dislocations, and these dislocation structures play crucial roles in determining the energy, dynamics and other properties of the grain boundaries \cite{ReadShockley1950,Sutton1995,Hirth2001}. As mentioned above, the classical grain boundary energy formula of Read and Shockley \cite{ReadShockley1950}  was obtained based on the dislocation structure of  low angle tilt boundaries in a simple cubic lattice that cancels the long-range elastic fields.
Vitek \cite{vitek1987} modified the grain boundary energy formula by including the interaction energy of intersecting dislocations in the dislocation structure of twist boundaries.
Rittner and Seidman \cite{Rittner1996} investigated $<110>$ symmetric tilt boundaries in fcc metals with low stacking fault energy by atomistic simulations, and developed a dislocation model of grain boundary dissociation by stacking fault emission.
Recently, Dai \textit{et al.} \cite{Dai20131327,Dai2014162} studied the structure and energy of fcc (111) twist grain boundaries using atomic, generalized Peierls-Nabarro and analytical models for all twist angles. They showed that dislocation structures on the twist boundaries can be determined by a single dimensionless parameter with two extreme cases of a hexagonal network of perfect dislocations and triangular network of partial dislocations enclosing stacking faults. Based on these dislocation structures, the twist grain boundary energy was derived as a function of twist angle, including the effects of partial dissociation and the stacking fault energy.
A microscopic phase field model was proposed by Shen \textit{et al.} \cite{Shen2014125} to describe the structures and energies of twist boundaries.
 Using discrete lattice sampling, their grain boundary energies agreed with the Read-Shockley model for low angle grain boundaries as well as the deep cusps for high angle special boundaries. Winther \textit{et al.} \cite{Winther2013,Winther2014} used discrete dislocation dynamics model to explain the dislocation networks in the deformation-induced grain boundaries aligned with slip planes in Al observed in experiments. They focused on the formation of grain boundaries by dislocation glide.
Lim \textit{et al.} \cite{Lim20095013,Lim20121395}
showed that the mobility of low angle grain boundaries under applied stress is determined by the constituent dislocation structures and their rearrangements by discrete dislocation dynamics simulations.
Wu and Voorhees \cite{Wu2012407} used the phase field crystal model to simulate the dynamic of a two-dimensional circular grain and observed motion and reaction of the constituent dislocations as the low angle grain shrinks. There are also models in the literature for the dislocation structures of heterogeneous interfaces \cite{Quek2011,Hirth2013,WangJian2013,Demkowicz2013,WangJianSciRep2013,Salehinia2014,Salehinia2015,Shao2015} and grain boundaries in hcp crystals \cite{Wang1-2012,Wang2-2012,WangJianSciRep2016}.

These recent works employed the discrete dislocation models or atomistic models. Although these models are able to provide detailed information on the dislocation or atomistic structures of individual grain boundaries, continuum model is desired for energetic and dynamics of grain boundaries at larger length scales.
For low angle grain boundaries, their energetic and dynamical properties  depend strongly on the dislocation structure, for example the grain boundary energy anisotropy \cite{Rohrer2011}. Except for the atomistic calculations by Olmsted \textit{et al.} \cite{Olmsted20093694,Holm2010905},
the available works that explore the dependence of grain boundary energy on all five DOFs by Saylor \textit{et al.} \cite{Saylor20033675} and Bulatov \textit{et al.} \cite{Bulatov2014161} are mainly based on energy reconstruction or interpolation and  do not directly depend on the dislocation microstructure of grain boundaries except for some special grain boundaries. Moreover, although it is well-known that the dislocation structure of a low angle grain boundary should satisfy the Frank's formula \cite{Frank1950,Bilby1955,Hirth2001}, this formula in general is not able to uniquely determine the dislocation structure. For example, there are six possible Burgers vectors  in an fcc crystal (neglecting the sign), leading to a total of twelve unknowns (two unknowns of orientation and interdislocation distance for the dislocation distribution of each Burgers vector). While in general the Frank's formula gives at most six linearly independent equations for a planar grain boundary. In the available discrete dislocation dynamics based works on the dislocation structures of low angle grain boundaries \cite{Winther2013,Winther2014} or heterogeneous interfaces \cite{Quek2011,Hirth2013,WangJian2013,Demkowicz2013} (for which a similar equation holds \cite{Knowles1982}), two or three prescribed Burgers vectors informed by experimental observations or atomistic simulations were adopted. There is no method available in the literature that is able to seek the dislocation structures over dislocations with all possible Burgers vectors, to the best of our knowledge.

In this paper,  we present a continuum model to compute the  energy of low angle grain boundaries for any given DOFs (arbitrary rotation axis, rotation angle and boundary plane orientation).  In our continuum model, we minimize the grain boundary energy associated with the dislocation structure subject to the constraint of Frank's formula with all possible Burgers vectors. This minimization problem is solved by the penalty method.
This model is based on the continuum framework for low angle grain boundaries \cite{Zhu2014175}, which gives total elastic energy including the long-range energy and local energy of dislocations in terms of dislocation densities on the grain boundary, and the latter is the grain boundary energy when a planar grain boundary is at equilibrium.
 Our continuum model can be considered a generalization of the classical Read-Shockley model, where a  closed-form energy formula was obtained for the special cases of tilt boundaries in a simple cubic lattice for which the Frank's formula is able to determine a unique dislocation structure on a boundary.

We use our continuum model to systematically study the energy of $<111>$ low angle grain boundaries in fcc Al with  any boundary plane orientation and all six possible Burgers vectors. Comparisons with results of atomistic simulations show that our continuum model is able to give excellent predictions of the energy and dislocation densities of low angle grain boundaries. We also study the energy of low angle grain boundaries in fcc Al with varying rotation axis while the remaining degrees of freedom are fixed, in which dislocations with all the six Burgers vectors are involved.

    Our continuum model is based on the representation of dislocation distributions by the dislocation density potential functions proposed in \cite{Zhu2014175}, which enables the generalization to curved grain boundaries. With modifications, our model can also apply to dislocation structures and energy of heterogeneous interfaces. These generalizations will be explored in the future work.

 This paper is organized as follows. In Sec.~2, we present our continuum simulation model. In Sec.~3, we validate our model by applications to the  energy of low angle tilt and twist boundaries and compare the results with those available in the literature.  In Sec.~4, using the developed continuum simulation model, we systematically study the energy of low angle grain boundaries in fcc Al with  $[111]$ rotation axis and any boundary plane orientation. The results are compared with atomistic simulation results. In Sec.~5, we further calculate the  energy of low angle grain boundaries in fcc Al with varying rotation axis while the remaining degrees of freedom are fixed. Possible generalizations of the present continuum model are discussed in Sec.~6.

\section{The Continuum Simulation Model}\label{method}

In this section,  we present a continuum simulation model to compute the energy of low angle grain boundaries based on densities of the constituent dislocations.  We
focus on planar low angle grain boundaries in this paper. The dislocation density potential functions proposed in Ref.~\cite{Zhu2014175} are adopted to describe the orientation-dependent dislocation densities on the grain boundaries. Assume the grain boundary  is the $xy$ plane. A
 dislocation density potential function $\eta$ is a scalar function defined on the $xy$ plane such that the constituent dislocations of the same Burgers vector $\mathbf b$ are given by the integer-valued contour lines of $\eta$: $\{\eta(x,y) = i, {\rm where}\ i  \ {\rm is \ an \ integer}\}$. The dislocation structure can be described in terms of $\nabla \eta$:
 the local dislocation line direction is $\mathbf{t} = (\nabla\eta /\|\nabla\eta\|)  \times \mathbf{n}$, where $\mathbf{n}$ is the unit normal vector of the grain boundary (which is in the $+z$ direction here), and  the inter-dislocation distance is $D = 1/\|\nabla \eta\|$.
 Recall that in the classical dislocation model of grain boundaries \cite{Sutton1995,Hirth2001}, the dislocation structure on a grain boundary is described by the reciprocal vector $\mathbf N$ that is lying in the boundary and perpendicular to the dislocation and has length $N=1/D$. Using our dislocation density potential function representation, $\mathbf N=\nabla \eta$. The advantage of our dislocation density potential function representation is that it also applies to curved dislocations on curved grain boundaries while maintaining the connectivity of dislocations.

 Assume that on the grain boundary, there are $J$ dislocation arrays represented by $\eta_j$, $j=1,2,\cdots,J$, corresponding to $J$ different Burgers vectors $\mathbf b_j$,  $j=1,2,\cdots,J$, respectively. We  solve the following problem for the dislocation structure and energy of the grain boundary:

\underline {Constrained Minimization Problem}:
\begin{eqnarray}
\text{minimize}& \nonumber\\
&& E = {\displaystyle \int_S \gamma_{\rm gb}  dS},\label{eqn:gb_energy}\\
{\rm with}&&\gamma_{\rm gb}={\displaystyle \sum_{j=1}^J  \frac{\mu(b^{(j)})^2}{4\pi(1-\nu)}\!\left[1-\nu\frac{(\nabla \eta_j\! \times \!\mathbf{n} \!\cdot\! \mathbf{b}^{(j)})^2}{(b^{(j)})^2 {\|\nabla \eta_j\|}^2}\right]\!\|\nabla \eta_j\| \log\! \frac{1}{r_g\sqrt{\|\nabla \eta_j\|^2+\epsilon}}},\label{eqn:gb_density}\\
\text{subject to}&&\nonumber\\
&&\mathbf{h}=\theta(\mathbf{V}\times \mathbf{a}) - {\displaystyle \sum_{j=1}^J} \mathbf{b}^{(j)}(\nabla\eta_j\cdot\mathbf{V})=\mathbf 0.\label{eqn:frank}
\end{eqnarray}
In this formulation, the grain boundary energy $E$ expressed in terms of densities of the constituent dislocations  \cite{Zhu2014175} is minimized subject to the constraint of Frank's formula $\mathbf h=\mathbf 0$.
 Here  $S$ is a period  on the grain boundary plane in terms of the dislocation structure on it (or the entire grain boundary if it is finite), $\gamma_{\rm gb}$ is the grain boundary energy density,
 $\mu$ is the shear modulus,  $\nu$ is the Poisson ratio,  $b^{(j)}$ is the length of the $j$-th Burgers vector, $r_g$  is a parameter depending on the size and energy of the dislocation core,
 $\epsilon$ is some small positive regularization parameter to avoid the numerical singularity when $\|\nabla \eta_j\|=0$,
 $\theta$ is the misorientation angle of the grain boundary, $\mathbf{a}$ is the unit vector along the rotation axis of the grain boundary, and $\mathbf V$ is any vector in the grain boundary plane.

  This formulation is based on the well-known fact that the dislocation structure of an equilibrium planar low angle grain boundary satisfies the Frank's formula \cite{Frank1950,Bilby1955,Hirth2001}, which is given in Eq.~\eqref{eqn:frank}.
As discussed in the introduction that Frank's formula in general is not able to uniquely determine the dislocation structure, we minimize the grain boundary energy over all the dislocation structures that satisfy the Frank's formula. The grain boundary energy formula in Eqs.~\eqref{eqn:gb_energy} and \eqref{eqn:gb_density}  was derived in Ref.~\cite{Zhu2014175}. (See also the elastic energy expression for dislocation densities in the bulk \cite{nelson1981,rickman1997}.) Note that under the constraint of Frank's formula, the long-range elastic energy of the constituent dislocations vanishes and the grain boundary energy equals the dislocation line energy \cite{Frank1950,Bilby1955,Hirth2001,Zhu2014175}.

Numerically, the constrained minimization problem is solved by the penalty method \cite{Chong2013}, in which  the problem is approximated by the following unconstrained minimization problem.

\underline {Unconstrained Minimization Problem}:
\begin{equation}\label{penalty}
\text{minimize } Q = {\displaystyle \int_S \left(\gamma_{\rm gb}   +\frac{1}{2}\alpha \|\mathbf h\|^2\right) dS},
\end{equation}
where $\alpha >0$ with large value is the penalty parameter. It has been shown that as $\alpha\rightarrow+\infty$, the solution of this unconstrained minimization problem converges to the solution of the constrained minimization problem \cite{Chong2013}.
This unconstrained problem with curved dislocations and nonuniform dislocation densities is still very challenging to solve due to the nonconvexity of the grain boundary energy. We make a further simplification by considering uniform distributions of straight dislocations on the grain boundary. In this case, each $\nabla \eta_j=(\eta_{jx},\eta_{jy})$ is a constant vector, and the problem is reduced to minimize $q=\gamma_{\rm gb}+\alpha \|\mathbf h\|^2/2$. This unconstrained problem can be solved by the  gradient minimization method  as follows.

Assume that in the current coordinate system where the grain boundary plane is the $xy$ plane and its normal direction is the $z$ direction, the Burgers vectors are  $\mathbf{b}^{(j)}=
(s_{j1},s_{j2},s_{j3})b^{(j)}$, $j=1,2,\cdots,J$, and the rotation axis is $\mathbf{a}=(a_1,a_2,a_3)$. The Frank's formula in Eq.~\eqref{eqn:frank} holds if and only if it holds for the two basis vectors of the $xy$ plane: $\mathbf{V}=\mathbf{V_1}=(1,0,0)$ and $\mathbf{V}=\mathbf{V_2}=(0,1,0)$. Using these expressions, gradient minimization of the unconstrained problem in Eq.~\eqref{penalty} with respect to variables $\eta_{jx}$,  $\eta_{jy}$, $j=1,2,\cdots,J$,
leads to the following evolution equations:
\begin{eqnarray}\label{}
&&(\eta_{jx})_t =-\left( \frac{\partial \gamma_{\rm gb}}{\partial \eta_{jx}} + \alpha\frac{\partial p}{\partial \eta_{jx}}\right),\vspace{1ex}\label{eqn:evolution0}\\
&&(\eta_{jy})_t =-\left( \frac{\partial \gamma_{\rm gb}}{\partial \eta_{jy}} + \alpha\frac{\partial p}{\partial \eta_{jy}}\right),\label{eqn:evolution1}
\end{eqnarray}
for $j=1,2,\cdots,J$, where  $p=\|\mathbf h\|^2/2$, and $\mathbf h=(h_1,h_2,\cdots,h_6)$ with
 $h_1 = - \sum_{j=1}^J b^{(j)}s_{j1} \eta_{jx}$,
 $h_2 = -\sum_{j=1}^J b^{(j)}s_{j2} \eta_{jx} - \theta a_3$,
 $h_3 = -\sum_{j=1}^J b^{(j)}s_{j3} \eta_{jx} +\theta a_2$,
 $h_4 = -\sum_{j=1}^J b^{(j)}s_{j1} \eta_{jy} + \theta a_3$,
 $h_5 = -\sum_{j=1}^J b^{(j)}s_{j2} \eta_{jy}$, and
 $h_6 = -\sum_{j=1}^J b^{(j)}s_{j3} \eta_{jy} - \theta a_1$.

  The parameter $r_g$ in this formulation depends on the dislocation core energy $E_c$ and the dislocation core size $r_0$.  In principle it can be obtained from the derivation of the continuum model from the discrete dislocation model \cite{Zhu2014175}. However, such a formula will be very complicated and not practical to use (see an example in Ref.~\cite{Zhu2011}) Analytical formulas for $r_g$ in terms of $E_c$ and $r_0$ are available for some special low angle grain boundaries such as symmetric tilt and pure twist boundaries \cite{ReadShockley1950,vitek1987}, for which grain boundary energy formulas are available. (See an example of formula of $r_g$ in terms of $E_c$ and $r_0$ for a symmetric tilt boundary given at the end of Sec.~3.1.)
 The dislocation core energy $E_c$ and the dislocation core size $r_0$ in these available grain boundary energy formulas are commonly fitted from atomistic simulation data. Here we choose to directly fit the  parameter $r_g$ from atomistic results for these special grain boundaries (see Sec.~3), and  then extend these values of $r_g$ to all grain boundary orientations by an interpolation (see Sec.~4).

We consider grain boundaries in fcc crystals, in which there are $J=6$  Burgers vectors of the $<110>$ type with the same length $b$ (neglecting the sign). As an example, when the  grain boundary is the $(111)$ plane, we can choose the directions $[\bar{1}10]$, $ [\bar{1}\bar{1}2]$ and $[111]$ to be  the $x,y,z$ directions, respectively, in our simulations. In this coordinate system, the six Burgers vectors are
$\mathbf{b}^{(1)} = \left(1,0,0\right)b$, $\mathbf{b}^{(2)} = \left(\frac{1}{2},\frac{\sqrt{3}}{2},0\right)b$, $\mathbf{b}^{(3)} = \left(\frac{1}{2},-\frac{\sqrt{3}}{2},0\right)b$, $\mathbf{b}^{(4)} = \left(0,\frac{\sqrt{3}}{3},\frac{\sqrt{6}}{3}\right)b$, $\mathbf{b}^{(5)} = \left(\frac{1}{2},\frac{\sqrt{3}}{6},-\frac{\sqrt{6}}{3}\right)b$, and $\mathbf{b}^{(6)} = \left(-\frac{1}{2},\frac{\sqrt{3}}{6},-\frac{\sqrt{6}}{3}\right)b$.  Recall that we always choose the normal direction of the grain boundary to be the $z$ axis in our simulation. The coordinates of these vectors in this computational coordinate system can be calculated by linear transformation from their crystallographic coordinates.
(Remark on linear transformation: Assume that $\mathbf l_1$, $\mathbf l_2$ and $\mathbf l_3$ are the three orthogonal unit vectors (column vectors) that are transformed into the $x$, $y$ and $z$ axes, respectively. A vector $\mathbf l$ in the old coordinate system is transformed into the vector $\mathbf l_{\rm new}=\mathbf R\mathbf l$ in the new coordinate system, where $\mathbf R=(\mathbf l_1,\mathbf l_2,\mathbf l_3)^T$  is the transformation matrix and the superscript $T$ means matrix transpose.) In the simulations in this paper, we focus on fcc Al, whose Burgers vectors  have length $b=0.286nm$. The regularization parameter $\epsilon$ in Eq.~\eqref{eqn:gb_density} is chosen to be $8\times 10^{-7}b^{-2}$.

\section{Pure tilt and twist boundaries}

 In this section, we apply our continuum simulation model to the energy of
  low angle tilt and twist boundaries. The rotation axis $\mathbf{a}$ of the boundaries is fixed to be  in the $[111]$ direction. These simulations serve as examples to validate our model. Moreover, through these simulations, we calibrate the values of $r_g$  for the  pure tilt and pure twist boundaries
by the  available results of energies of these boundaries by using molecular statics (MS) or molecular dynamics (MD) simulations. These  values of $r_g$ will be extended to all grain boundary orientations by an interpolation in the next section. We use the EAM potential for Al developed by Mishin \textit{et al.}  \cite{Mishin1999} and the LAMMPS code \cite{Plimpton1995} in the MS simulations.

Recall that in the continuum model, the locations of dislocations are the integer-value contour lines of the dislocation density potential functions $\eta_j$, $j=1,2,\cdots,6$, and in the simulations in this paper, $\eta_j(x,y) = \eta_{jx} x + \eta_{jy} y$, where $\eta_{jx}$ and $\eta_{jy}$ are constants and are obtained by solving the minimization problem in Eq.~\eqref{penalty} (that is, the evolution equations in Eqs.~\eqref{eqn:evolution0} and \eqref{eqn:evolution1}). We start from $\eta_{jx}=\eta_{jy}=0$, $j=1,2,\cdots,6$, when solving these equations. We choose a large value for the penalty parameter $\alpha$ and further increases of its value give only negligible changes in the converged dislocation structures.

\subsection{Tilt boundaries}

 We first consider the low angle $[111]$ symmetric tilt boundaries whose boundary plane normal is in the   $[\bar{1}10]$ direction. In the simulations, we choose the directions $[\bar{1}\bar{1}\bar{1}]$,  $[\bar{1}\bar{1}2]$, and $[\bar{1}10]$ to be  the $x,y,z$ directions, respectively. In this coordinate system, the six Burgers vectors are
$\mathbf{b}^{(1)} = \left(0,0,1\right)b$, $\mathbf{b}^{(2)} = \left(0,\frac{\sqrt{3}}{2},\frac{1}{2}\right)b$, $\mathbf{b}^{(3)} = \left(0,-\frac{\sqrt{3}}{2},\frac{1}{2}\right)b$, $\mathbf{b}^{(4)} = \left(-\frac{\sqrt{6}}{3},\frac{\sqrt{3}}{3},0\right)b$, $\mathbf{b}^{(5)} = \left(\frac{\sqrt{6}}{3},\frac{\sqrt{3}}{6},\frac{1}{2}\right)b$, $\mathbf{b}^{(6)} = \left(\frac{\sqrt{6}}{3},\frac{\sqrt{3}}{6},-\frac{1}{2}\right)b$, and the rotation axis is $\mathbf{a}=(-1,0,0)$.

\begin{figure}[htbp]
\centering
    \includegraphics[width=0.55\linewidth]{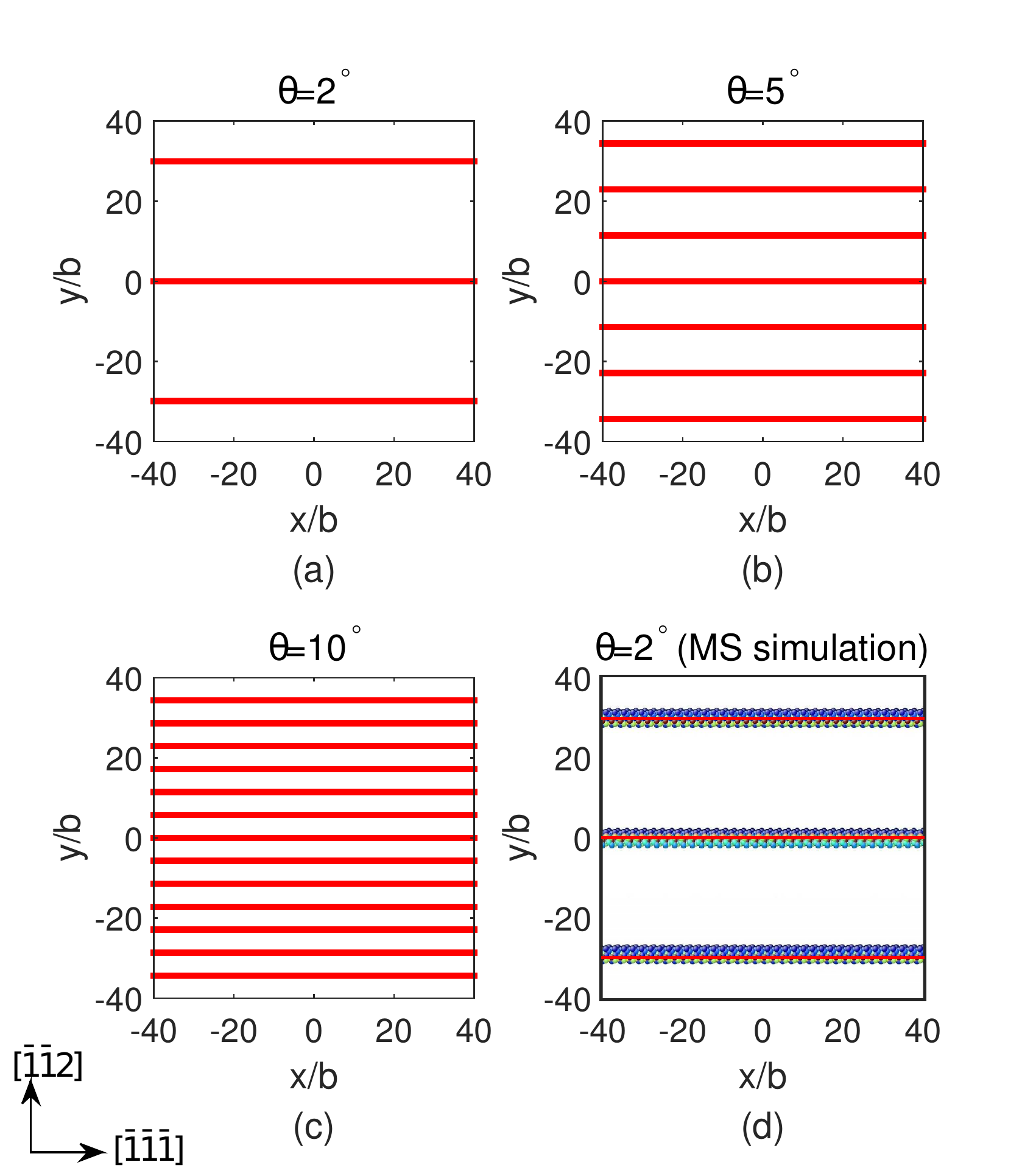}
    \caption{(a)-(c) Dislocation structure of low angle $[111]$ symmetric tilt boundaries with different misorientation angles  calculated using our continuum model. The dislocation structure consists of only the array of $\mathbf{b}^{(1)}$-dislocations. (d) Dislocation locations compared with the MS results  for misorientation angle $\theta=2^\circ$. Blue atoms: MS results. Red lines: continuum simulation. }
    \label{eta_010tilt}
\end{figure}

Simulation results  show that the dislocation structure of such a tilt boundary consists of only the array of dislocations with the Burgers vector $\mathbf{b}^{(1)}$ represented by the dislocation density potential function $\eta_1$. These dislocations are lying parallel to the $x$ axis correspond to the $[\bar{1}\bar{1}\bar{1}]$ direction. Recall that the Burgers vector $\mathbf{b}^{(1)}$ is in the $z$ direction corresponding to the $[\bar{1}10]$ direction. Dislocations with other Burgers vectors do not appear in the converged dislocation structure. That is, $\eta_2, \eta_3, \cdots, \eta_6$ converge to $0$ during the evolution.
  Some obtained dislocation structures for different misorientation angles are shown in Fig.~\ref{eta_010tilt}(a)-(c). It can be seen that the dislocation density increases as the misorientation angle $\theta$ increases. We make quantitative comparisons between the
dislocation density obtained by using our continuum model and the theoretical value $1/D=\theta/b$
using the dislocation model \cite{ReadShockley1950,Sutton1995,Hirth2001}
as  shown in the Table~\ref{TiltDensityComp}, where $D$ is the inter-dislocation distance. Excellent agreement can be seen from the comparisons.  We have also performed MS simulations for these tilt boundaries. Excellent agreement is also found between the results using the continuum simulation model and the MS model, see an example in Fig.~\ref{eta_010tilt}(d).

\begin{table}[htbp]
\centering
\caption{Dislocation Density on the $[111]$ Symmetric  Tilt Boundaries (unit: $b^{-1}$).}
\vspace{1ex}
\begin{tabular}{ccccc}
\hline
$\theta$ & $3^\circ$& $5^\circ$& $7.5^\circ$& $10^\circ$\\
\hline
\text{Our simulation}&0.0522&0.0872&0.1745&0.3490 \\
\hline
\text{Theoretical value  $1/D=\theta/b$}&0.0524&0.0873&0.1745&0.3491\\
\hline
\end{tabular}\\
\label{TiltDensityComp}
\end{table}

\begin{figure}[htbp]
 \centering
    \includegraphics[width=0.5\linewidth]{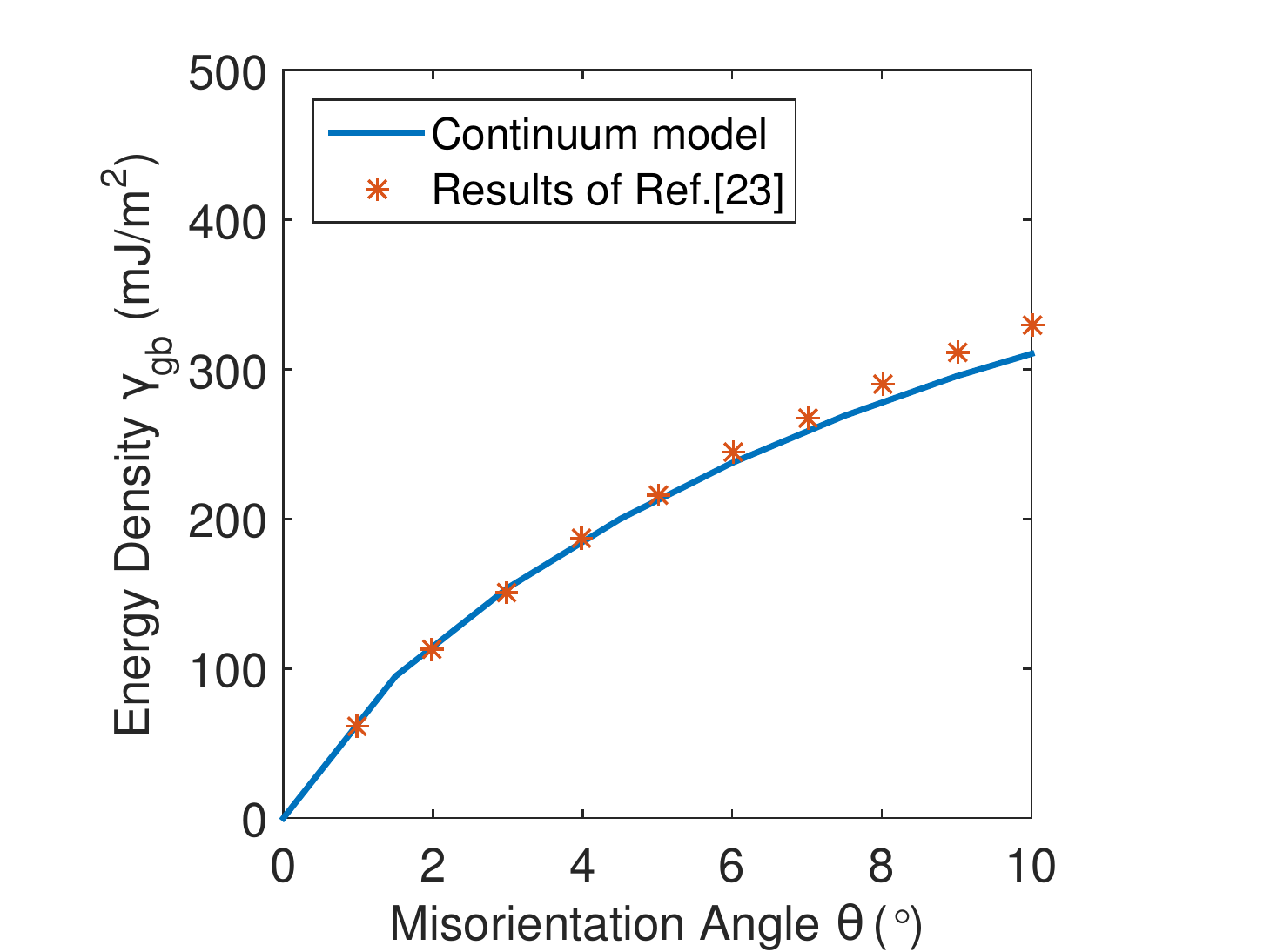}
    \caption{Grain boundary energy density  of $[111]$ symmetric tilt boundaries in the low angle regime calculated by Eq.~\eqref{eqn:gb_density} using the obtained dislocation structure (described by $\eta_j$, $j=1,2,\cdots,6$), and
    comparison with the available results in Ref.~\cite{Bulatov2014161}.}
    \label{(010)tilt_energy}
\end{figure}

With the obtained dislocation structure, the grain boundary energy density can be calculated by Eq.~\eqref{eqn:gb_density}. The calculated energy density of these tilt boundaries as a function of the misorientation angle $\theta$ within the low angle regime is plotted in Fig.~\ref{(010)tilt_energy}. We also compare our results with those in Ref.~\cite{Bulatov2014161} obtained by fitting the MD data in Fig.~\ref{(010)tilt_energy}. Excellent agreement can be found  if we set $r_g=0.85b$  in Eq.~\eqref{eqn:gb_density}. Note that in the simulations, the converged dislocation structure is not sensitive to the value of $r_g$. In the calibration process to determine the value of $r_g$, we start with a prescribed value $r_g=b$ to calculate the dislocation structure and then fit the value of $r_g$ in the energy formula by available data; with this more accurate value of $r_g$, the dislocation structure and energy are calculated again using our continuum model.

  Note that using the classical dislocation model of grain boundaries \cite{ReadShockley1950,Sutton1995,Hirth2001}, the energy of this symmetric tilt boundary   is $\theta(E_c/b+(\mu b/4\pi(1-\nu))\log (eb/2\pi r_0\theta))$, where $E_c$ is recalled to be the dislocation core energy and $r_0$ the dislocation core size. For this dislocation structure, our grain boundary energy in Eq.~\eqref{eqn:gb_density} is $\gamma_{\rm gb}= (\mu b \theta/4\pi(1-\nu))\log (b/r_g\theta)$. These give the analytical formula of $r_g$ in this case: $\log(r_g/b)=-E_c/(\mu b^2 /4\pi(1-\nu))+ \log(2\pi r_0/eb)$. The value of $r_g$ in this case can also be calculated using this formula, if the values of $E_c$ and $r_0$ are available.

\subsection{Twist boundaries}

 We then consider the low angle $[111]$ twist boundaries. The coordinate system in our simulations and the coordinates of the Burgers vectors  have been given at the end of Sec.~2, and the rotation axis is $\mathbf a=(0,0,1)$ in this coordinate system.

\begin{figure}[htbp]
\centering
    \includegraphics[width=0.75\linewidth]{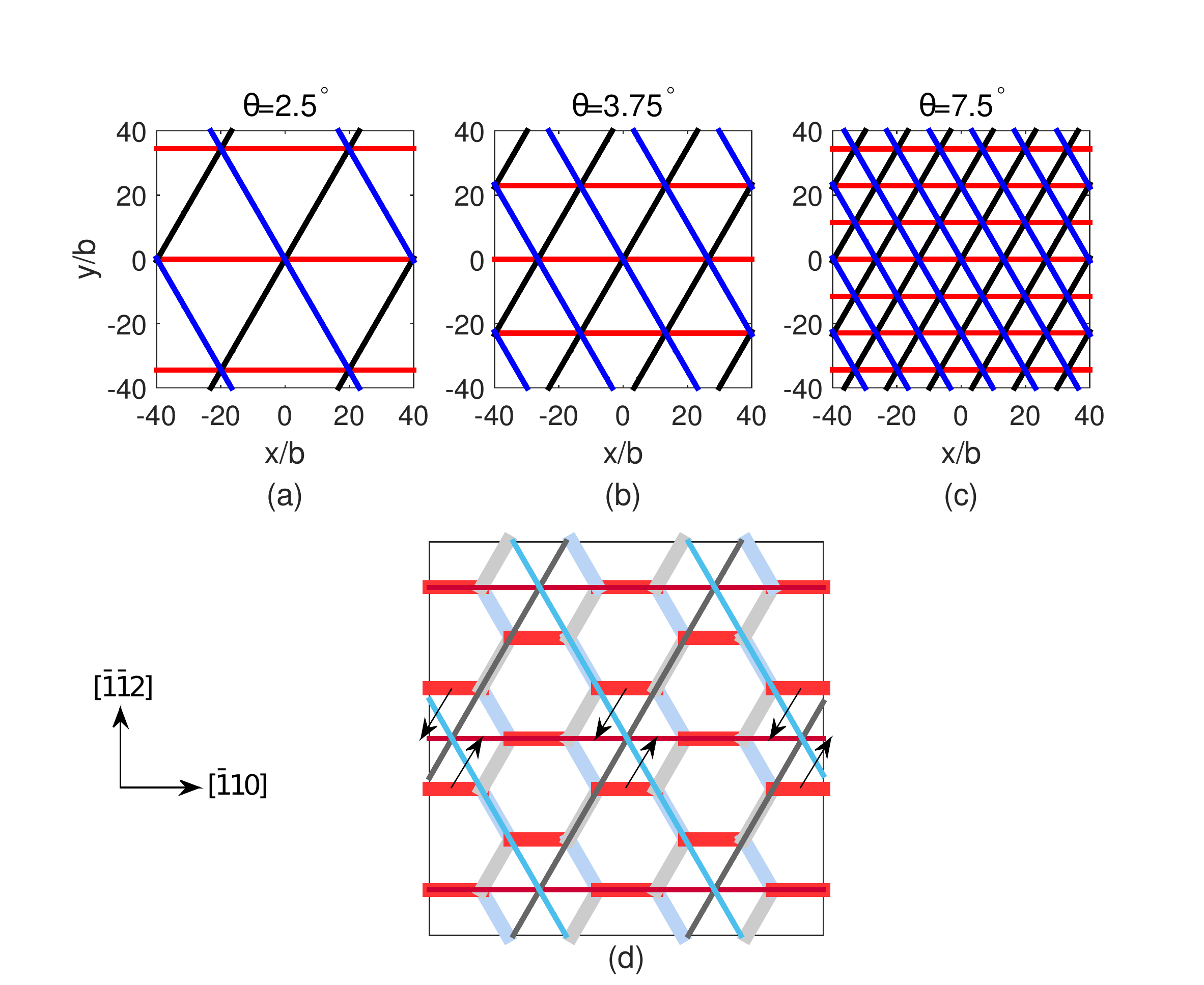}
    \caption{(a)-(c) Dislocation structure of low angle $[111]$ twist boundaries with different misorientation angles  calculated using our continuum model. The dislocation structure consists of  a triangular network of three arrays of screw dislocations with Burgers vectors  $\mathbf{b}^{(1)}$, $\mathbf{b}^{(2)}$, and $\mathbf{b}^{(3)}$, shown by red, black, and blue lines, respectively.  (d) Comparison of the dislocation structure by our continuum simulation and the exact dislocation structure of hexagonal network by the discrete dislocation model and  MS simulations. The dislocations of each of the three Burgers vectors in the exact hexagonal network are shown by thick line segments with the same but lighter color as those in our simulation results. In our continuum model, dislocations are straight lines that approximate the dislocation segments in the exact dislocation structure with same dislocation densities. }
    \label{eta_100twist}
\end{figure}

Simulation results using our continuum model show that the dislocation structure of such a twist boundary consists of  a triangular network of three arrays of screw dislocations with  Burgers vectors  $\mathbf{b}^{(1)}$, $\mathbf{b}^{(2)}$, and $\mathbf{b}^{(3)}$, see Fig.~\ref{eta_100twist}(a)-(c).  Recall that in our continuum simulation model, the grain boundary dislocation structure is approximated by densities of straight dislocations.
The dislocation structure obtained by our simulation is an approximation to   the exact dislocation structure of hexagonal network obtained by the discrete dislocation model \cite{Sutton1995,Hirth2001} and MS simulations \cite{Dai2014162} (without considering dislocation partial dissociation) in which the constituent dislocations are not straight.
 This approximation  is demonstrated in Fig.~\ref{eta_100twist}(d) by an example of comparison: The length of the red, horizontal dislocation line in our continuum model shown in the middle of Fig.~\ref{eta_100twist}(d) approximates the total length of the red dislocation segments in the exact structure on or near this straight line.
The purpose of our continuum model is to provide good approximations to the dislocation densities and grain boundary energy of the exact dislocation structure, as will be examined below.

\begin{table}[htbp]
\centering
\caption{Density of $\mathbf{b}^{(1)}$-Dislocations on the  $[111]$ Twist Boundaries (unit: $b^{-1}$).}
\vspace{1ex}
\begin{tabular}{ccccc}
\hline
$\theta$ & $2.5^\circ$& $3.75^\circ$& $5^\circ$& $7.5^\circ$\\
\hline
\text{Our simulation}&0.0291&0.0436&0.0582&0.0873 \\
\hline
\text{Theoretical value}&0.0282&0.0424&0.0565&0.0847\\
\hline
\end{tabular}\\
\label{TwistDensityComp}
\end{table}

\begin{figure}[htbp]
 \centering
    \includegraphics[width=0.5\linewidth]{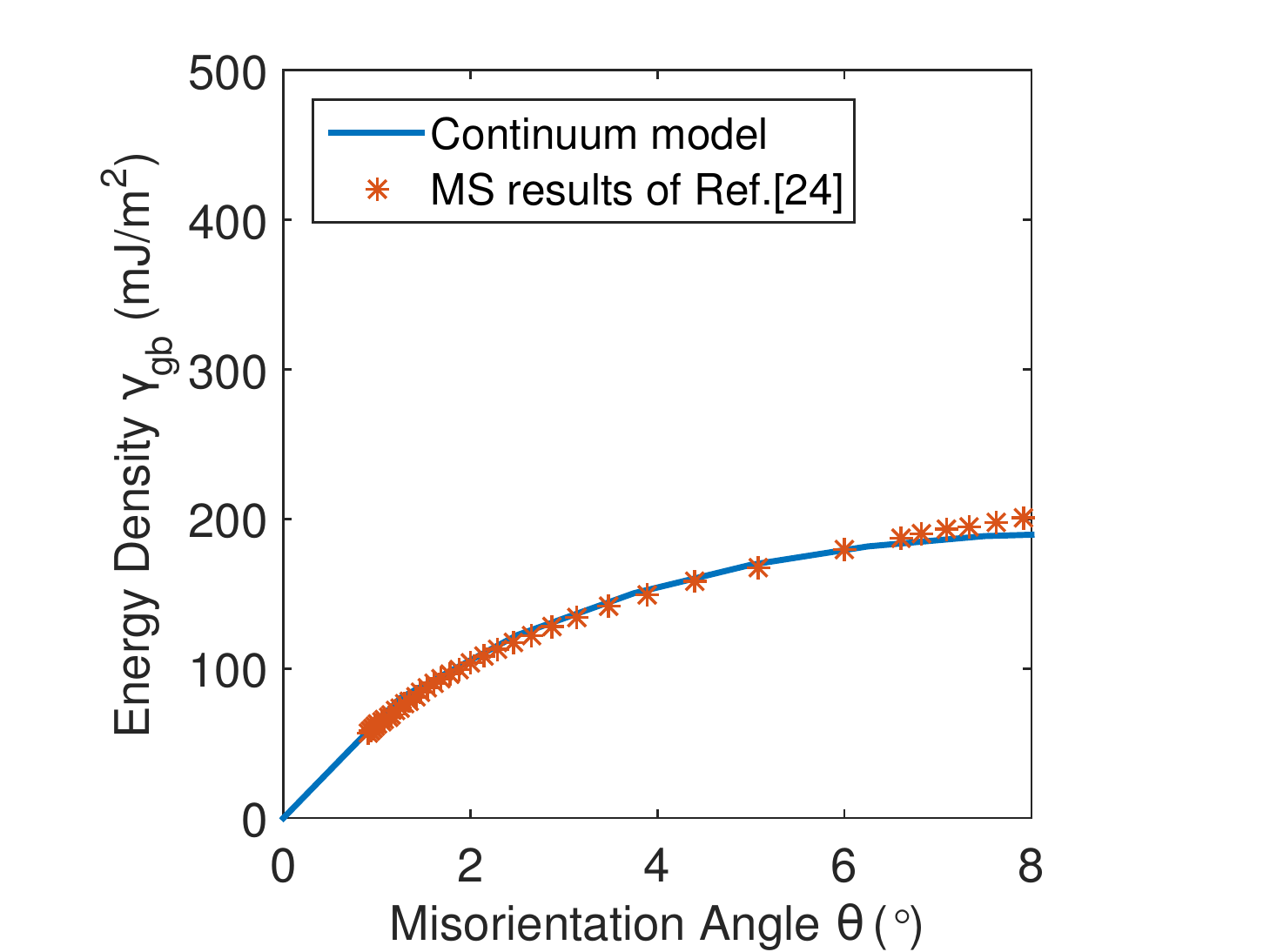}
    \caption{Grain boundary energy density  of $[111]$ twist boundaries in the low angle regime calculated by using our continuum model and
    comparison with the MS simulation results in Ref.~\cite{Dai2014162}.}
    \label{(100)twist_energy}
\end{figure}

Table~\ref{TwistDensityComp} shows the comparison of the density of $\mathbf{b}^{(1)}$-dislocations calculated using our continuum model and the theoretical value in the exact dislocation structure \cite{Sutton1995,Hirth2001} on these  $[111]$ twist boundaries. The calculated densities of $\mathbf{b}^{(2)}$ and $\mathbf{b}^{(3)}$-dislocations are almost identical to those of $\mathbf{b}^{(1)}$-dislocations,  and their theoretical values are the same.
It can be seen from this table that our continuum simulation model gives excellent approximations to the theoretical values of  dislocation densities, even though the simplified structure of straight dislocations is used in the continuum model.

Fig.~\ref{(100)twist_energy} shows
 the energy density  of these twist boundaries as a function of the misorientation angle $\theta$ within the low angle regime, calculated by using our continuum model. We calibrate the parameter $r_g$ in Eq.~\eqref{eqn:gb_density} for these twist boundaries by using the MS simulation results in Ref.~\cite{Dai2014162}. Excellent agreement can be reached  if we set $r_g=3.5b$, see  Fig.~\ref{(100)twist_energy}.

\section{$<111>$ grain boundaries with arbitrary boundary plane orientation}

In this section, using the developed continuum simulation model, we systematically study the energy of low angle grain boundaries with rotation axis in the $[111]$ direction and an arbitrary boundary plane orientation. The misorientation angle of the grain boundary is fixed to be $\theta=1.95^\circ$.
We also perform MS simulations to examine the  dislocation structure and energy of these grain boundaries.

\subsection{$<111>$ grain boundaries parallel to $[\bar{1}\bar{1}2]$ direction}

\begin{figure}[htbp]
 \centering
    \includegraphics[width=0.5\linewidth]{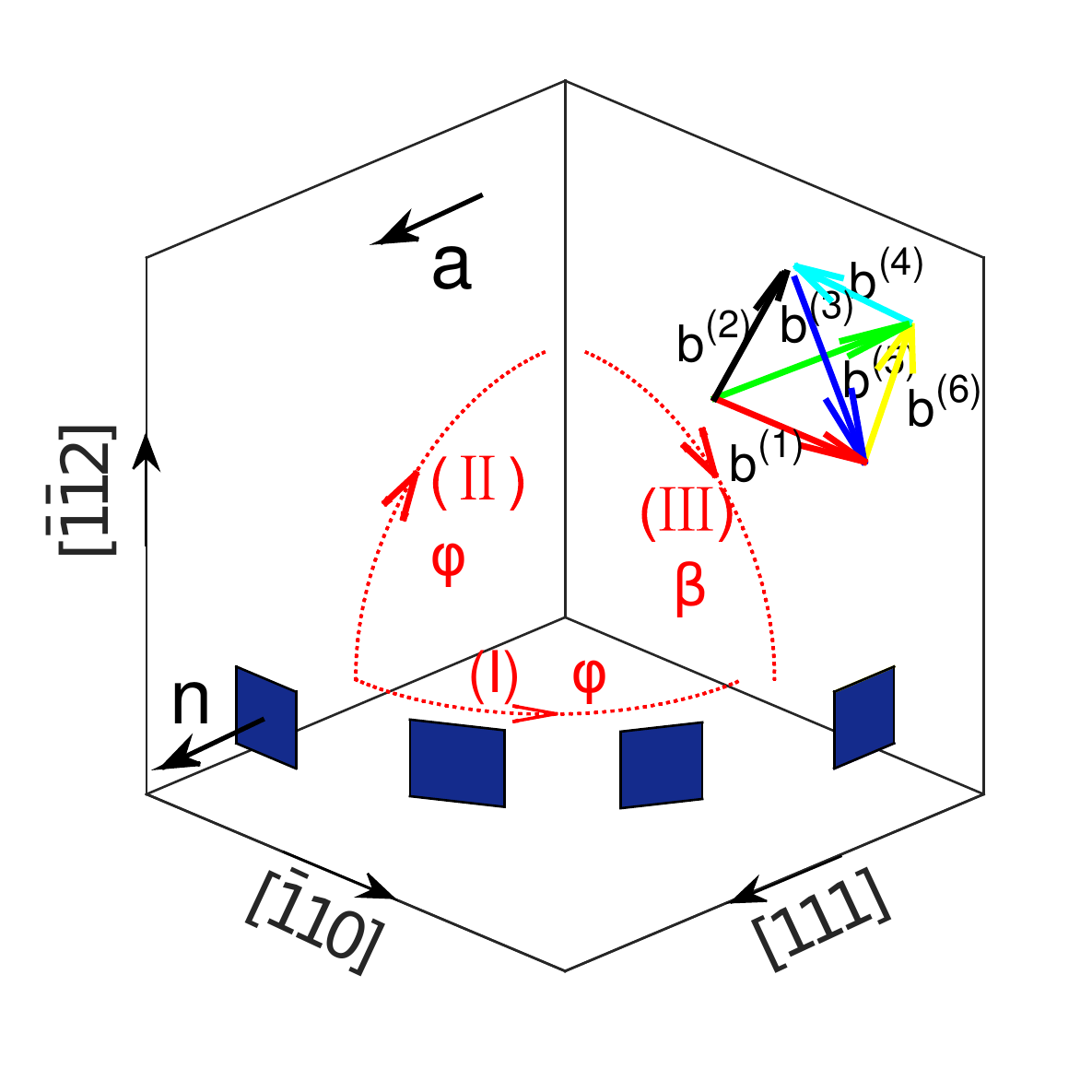}
    \caption{Grain boundaries parallel to the $[\bar{1}\bar{1}2]$ direction indexed by  the azimuthal angle $\phi$ with respect to  the $[111]$ direction. The rotation axis $\mathbf a$ is in the $[111]$ direction. The six Burgers vectors are $\mathbf b^{(j)}$, $j=1,2,\cdots,6$.}
    \label{structure21}
\end{figure}

We first consider the $[111]$ grain boundaries parallel to the $[\bar{1}\bar{1}2]$ direction. These grain boundaries can be indexed by an inclination angle $\phi$, which is the azimuthal angle of the normal vector of such a boundary with respect to  the $[111]$ direction, see path (I) in Fig.~\ref{structure21}. When $\phi=0^\circ$, the grain boundary is the $[111]$ pure twist boundary studied in Sec.~3.2; when $\phi=90^\circ$, it is the $[111]$ symmetric tilt boundary examined in Sec.~3.1.

 Recall that in the continuum formulation presented in Sec.~2, the grain boundary plane is always the $xy$ plane with normal in the $+z$ direction. The coordinates of the Burgers vectors $\mathbf b^{(j)}$, $j=1,2,\cdots,6$, and the rotation axis $\mathbf a$ in this simulation coordinate system can be calculated by linear transformation from their coordinates in the crystallographic coordinate system as reviewed in a remark in Sec.~2. More conveniently,  the coordinates of these vectors in the simulation coordinate system for such an angle-$\phi$ grain boundary can also be calculated directly from their simulation coordinates  for the twist boundary given at the end of Sec.~2, by the linear transformation of the angle-$\phi$ rotation along path (I). Note that in this case, $\mathbf l^{(1)}=(\cos\phi,0,-\sin\phi)^T$,  $\mathbf l^{(2)}=(0,1,0)^T$, and $\mathbf l^{(3)}=(\sin\phi,0,\cos\phi)^T$ in the transformation formulation there.

\begin{figure}[htbp]
    \includegraphics[width=0.9\linewidth]{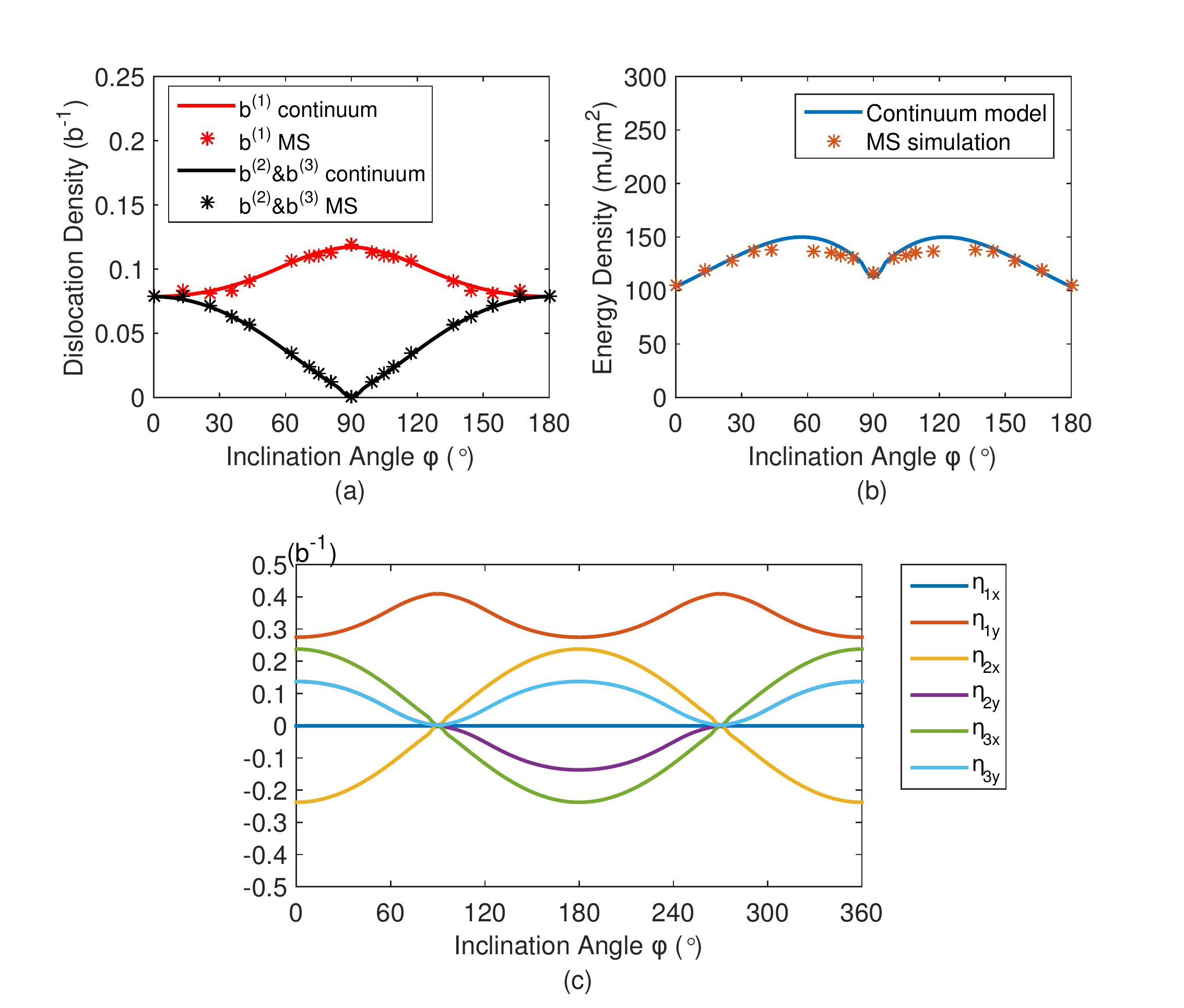}
    \caption{Dislocation densities (a) and grain boundary energy density (b) of the $[111]$ grain boundaries parallel to the $[\bar{1}\bar{1}2]$ direction indexed by the inclination angle $\phi$, computed by the continuum model and MS simulations. The misorientation angle is $\theta=1.95^\circ$. (c) The converged values of  $\eta_{jx}$ and $\eta_{jy}$ in our continuum model for the dislocation structure on these grain boundaries.  The $y$ axis in the continuum model is always in the $[\bar{1}\bar{1}2]$ direction. Note that $\eta_{jx}=\eta_{jy}=0$ for $j=4,5,6$, and the curve for $\eta_{2y}$ (with color purple) is identical to the curve for $\eta_{3y}$ (with color light blue) for $0^\circ\leq\phi\leq 90^\circ$ and $270^\circ\leq\phi\leq360^\circ$.}
    \label{DislocationDensity_xtoy}
\end{figure}

The values of  $\eta_{jx}$ and $\eta_{jy}$, $j=1,2,\cdots,6$, obtained  by solving the energy minimization problem in Eq.~\eqref{penalty} (that is, the evolution equations in Eqs.~\eqref{eqn:evolution0} and \eqref{eqn:evolution1}) are shown in Fig.~\ref{DislocationDensity_xtoy}(c) in terms of the inclination angle $\phi$.
  Recall that the locations of the $\mathbf b^{(j)}$-dislocations are integer-value contour lines of the dislocation density potential function $\eta_j(x,y)=\eta_{jx}x+\eta_{jy}y$, the density of these dislocations is $\rho_j=1/D_j=\|\nabla\eta_j\|=\sqrt{\eta_{jx}^2 + \eta_{jy}^2}$ where $D_j$ is the interdislocation distance in this array of dislocations, and the line direction of these dislocations is $\mathbf{t}_j = (\nabla\eta_j /\|\nabla\eta_j\|)  \times \mathbf{n}=(\eta_{jy},-\eta_{jx})/\sqrt{\eta_{jx}^2+\eta_{jx}^2}$.
  We can see in the results that for $j=1,2,3$, $\eta_{jx}$ and $\eta_{jy}$ vary with the inclination angle $\phi$, and $\eta_{jy}=\eta_{jz}=0$ for $j=4,5,6$.
 This means that the dislocation structure of these grain boundaries consists of dislocations with Burgers vectors $\mathbf b^{(1)}$, $\mathbf b^{(2)}$,  and $\mathbf b^{(3)}$, and no dislocations with Burgers vectors $\mathbf b^{(4)}$, $\mathbf b^{(5)}$, or $\mathbf b^{(6)}$ appear in the structure. The dislocation structure  is  symmetric with respect to $\phi=0^\circ$, $90^\circ$, $180^\circ$ and $270^\circ$ when the difference due to opposite line directions is neglected.
  The calculated densities of these dislocations and comparison with the MS results are shown in Fig.~\ref{DislocationDensity_xtoy}(a). It can be seen that our continuum model is able to give excellent approximations to the densities of dislocations on these grain boundaries.
The results of both models show that as $\phi$ varies from $0^\circ$ (pure twist) to $90^\circ$ (symmetric tilt), the density of $\mathbf b^{(1)}$-dislocations increases, and both densities of $\mathbf b^{(2)}$ and $\mathbf b^{(3)}$-dislocations   gradually decrease to $0$.

\begin{figure}[htbp]
    \includegraphics[width=1.0\linewidth]{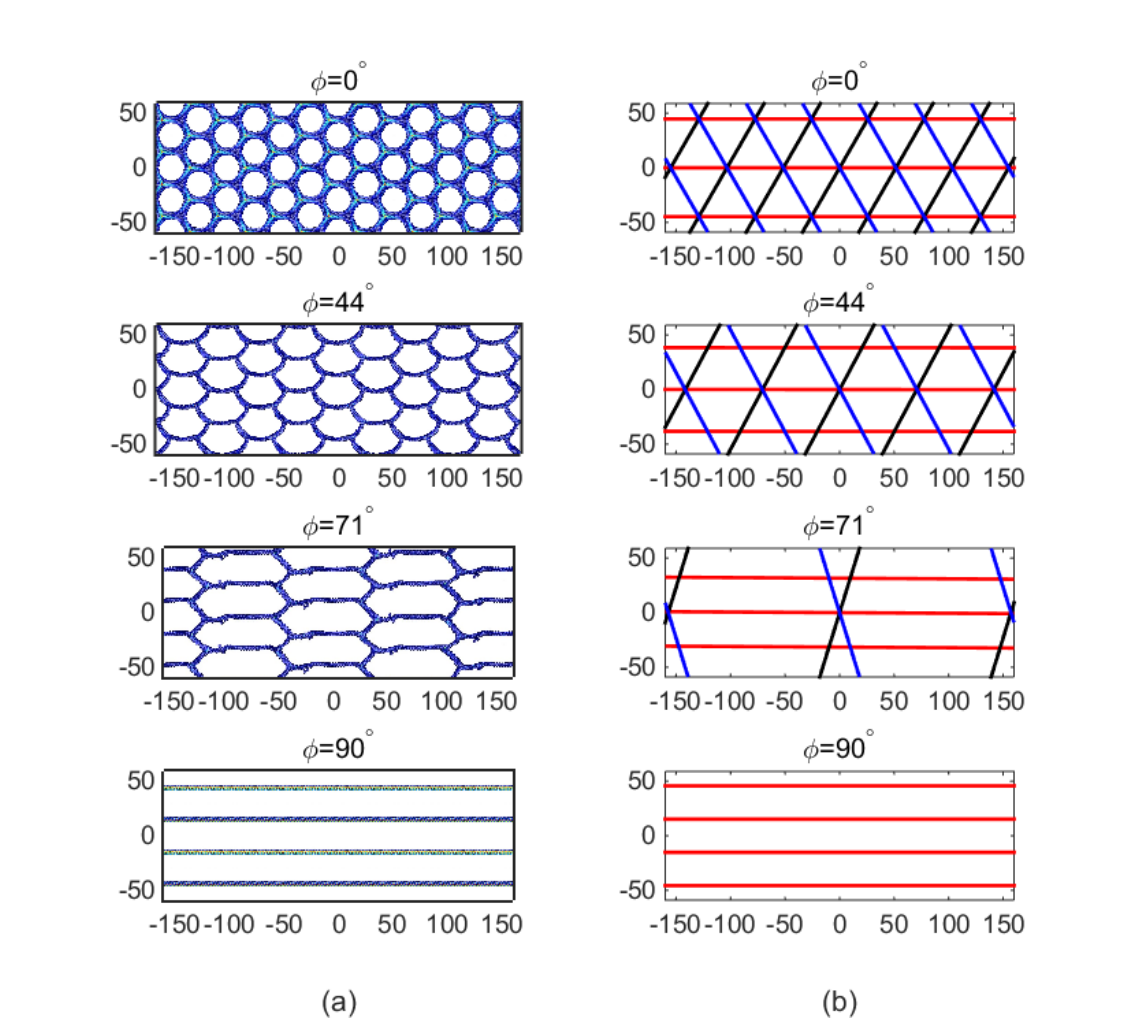}
    \caption{Dislocation structure of the $[111]$ grain boundaries parallel to the $[\bar{1}\bar{1}2]$ direction indexed by the inclination angle $\phi$, computed by (a) MS simulations and (b) the continuum model. The vertical  direction is the $[\bar{1}\bar{1}2]$ direction (the $y$ axis in the continuum model). The misorientation angle is $\theta=1.95^\circ$. As a continuum model, our method is able to provide accurate approximations to the dislocation densities and energy of these low angle grain boundaries, whereas not all the details of the dislocation structure on the levels of discrete dislocation dynamics and/or atomistic model are resolved in our continuum model due to the straight-dislocation approximation.}
    \label{eta_z}
\end{figure}

 Fig.~\ref{DislocationDensity_xtoy}(b) shows
 the energy density  of these grain  boundaries as a function of the inclination angle $\phi$  calculated by our continuum model. We calibrate the parameter $r_g$ in Eq.~\eqref{eqn:gb_density} for these grain boundaries by using our MS simulation results. As shown in Fig.~\ref{DislocationDensity_xtoy}(b), excellent agreement can be reached  if we set
 \begin{equation}
 r_g=3.5 e^{-1.4\sin\phi}b,
 \label{eqn:rg111}
 \end{equation}
where $\phi$ is recalled to be the  azimuthal angle of the grain boundary normal  with respect to  the $[111]$ rotation axis $\mathbf a$. This form of the formula of $r_g$  is chosen such that its contribution to the grain boundary energy in Eq.~\eqref{eqn:gb_density} is  $\log(r_g/b)=A+B\sin\phi$, where $A$ and $B$ are constants, and  $A=\log3.5$, $B=-1.4$ by fitting the MS results.
It can be seen from Fig.~\ref{DislocationDensity_xtoy}(b) that as the inclination angle $\phi$ varies from $0^\circ$ to $90^\circ$, i.e. the grain boundary varies from the pure twist to symmetric tilt following this path,
the grain boundary energy increases first and finally decreases, ending up with a higher energy of the symmetric tilt boundary than that of the pure twist boundary. Based on the obtained dislocation structure on these boundaries, as  $\phi$ increases from $0^\circ$ to $90^\circ$,
the density of $\mathbf b^{(1)}$-dislocations increases and the character of these dislocations changes from pure screw to pure edge; both of the changes increase the energy.
 Whereas along this path, the densities of $\mathbf b^{(2)}$-dislocations and $\mathbf b^{(3)}$-dislocations decrease and these dislocations have more and more edge character; the former change decreases the energy while the latter increases it. Combination of these effects leads to the energy behavior shown in Fig.~\ref{DislocationDensity_xtoy}(b).

Fig.~\ref{eta_z} shows the detailed dislocation structures on these grain boundaries using the continuum model (Fig.~\ref{eta_z}(b)) and MS simulations (Fig.~\ref{eta_z}(a)), from the boundary of $\phi=0$ which is the pure twist boundary discussed in Sec.~3.2 to the boundary of $\phi=90^\circ$ which is the symmetric tilt boundary discussed in Sec.~3.1. The dislocation structures of the grain boundaries with other values of $\phi$ are similar due to the symmetries   discussed above. We have shown that as a continuum model, our method is able to provide accurate approximations to the dislocation densities and energy of these low angle grain boundaries. It can also be seen from Fig.~\ref{eta_z} that the dislocation line directions obtained by our continuum model is also consistent with those by MS simulations. However, not all the details of the dislocation structure on the levels of discrete dislocation dynamics and/or atomistic model are resolved in our continuum model, mainly due to the straight-dislocation approximation.
  How to understand the approximation of dislocation densities in our continuum model  has been demonstrated in Fig.~\ref{eta_100twist}(d).
 When the constituent dislocations of the grain boundaries are all straight, our continuum model is able to give the exact dislocation structure given by MS simulation, as shown in the last row of images in Fig.~\ref{eta_z} for the symmetric tilt boundary. An interesting observation on the atomic details of these grain boundaries is that when $\phi$ is between $0^\circ$ and $90^\circ$, the grain boundaries obtained by MS simulations are not planar: some dislocation triple junctions move into the plane and the remaining ones move out of it. The influence of these atomic details on the grain boundary energy is negligible, whereas they may slightly modify the orientations of the constituent dislocations.

\subsection{$<111>$ grain boundaries parallel to $[\bar{1}10]$ direction}

In this subsection, we consider the $[111]$ grain boundaries parallel to the $[\bar{1}10]$ direction. These grain boundaries are also indexed by the azimuthal angle $\phi$ of the grain boundary normal with respect to  the $[111]$ direction, see path (II) in Fig.~\ref{structure21}. When $\phi=0^\circ$, the grain boundary is the $[111]$ pure twist boundary studied in Sec.~3.2; When $\phi=90^\circ$, it is  a $(\bar{1}\bar{1}2)[111]$ tilt boundary.

\begin{figure}[htbp]
\centering
    \includegraphics[width=0.9\linewidth]{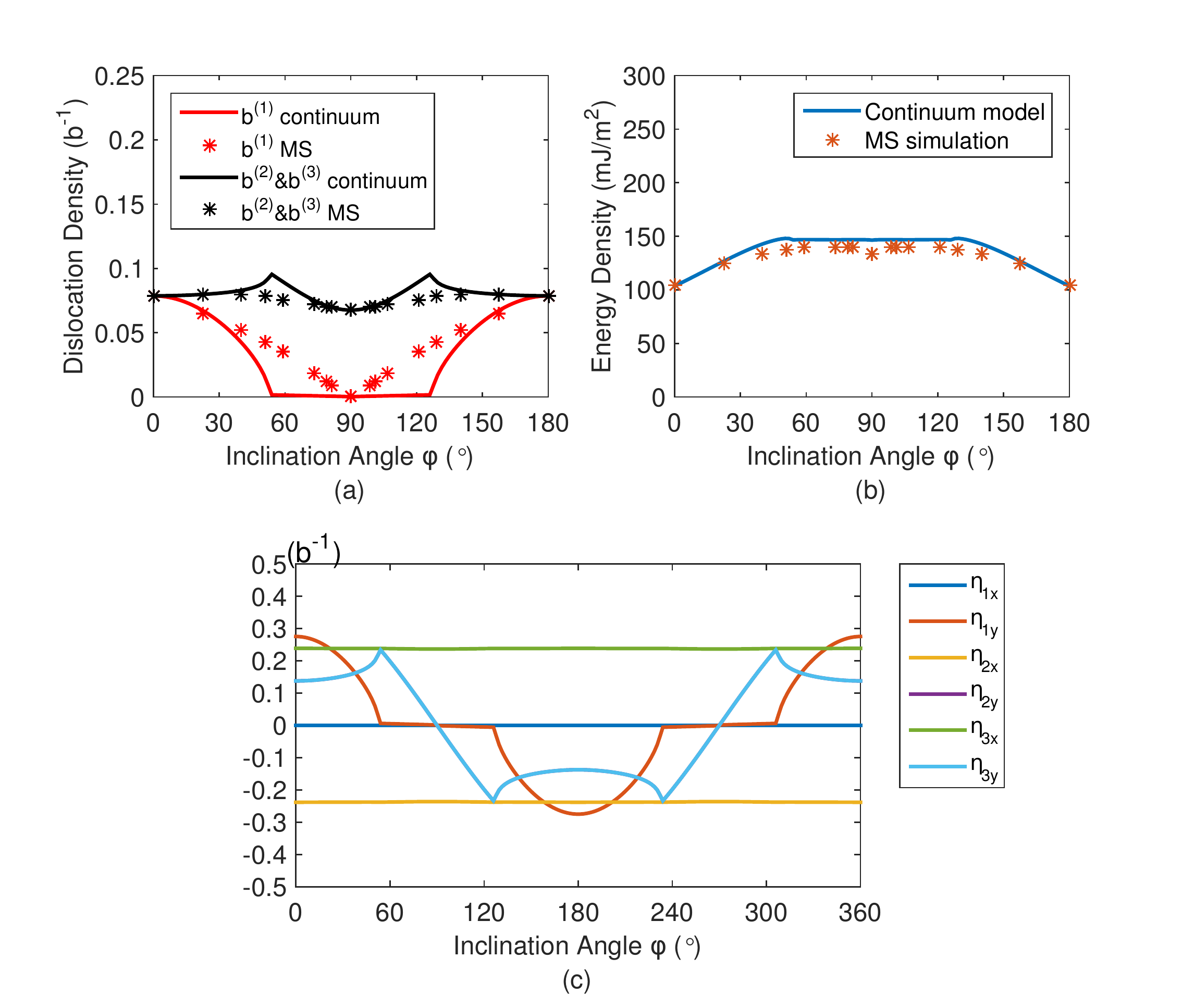}
    \caption{Dislocation densities (a) and grain boundary energy density (b) of the $[111]$ grain boundaries parallel to the $[\bar{1}10]$ direction indexed by the inclination angle $\phi$, computed by the continuum model and MS simulations. The misorientation angle is $\theta=1.95^\circ$. (c) The converged values of  $\eta_{jx}$ and $\eta_{jy}$ in our continuum model for the dislocation structure on these grain boundaries. The $x$ axis in the continuum model is always in the $[\bar{1}10]$ direction.  Note that  $\eta_{jx}=\eta_{jy}=0$ for $j=4,5,6$.}
    \label{DislocationDensity_xtoz}
\end{figure}

The values of  $\eta_{jx}$ and $\eta_{jy}$, $j=1,2,\cdots,6$, on these grain boundaries obtained  by using the continuum model are shown in Fig.~\ref{DislocationDensity_xtoz}(c) in terms of the inclination angle $\phi$.
As in the grain boundaries discussed in the previous subsection,  the dislocation structure of these grain boundaries also consists of dislocations with Burgers vectors $\mathbf b^{(1)}$, $\mathbf b^{(2)}$,  and $\mathbf b^{(3)}$. The dislocation structure  is  also symmetric with respect to $\phi=0^\circ$, $90^\circ$, $180^\circ$ and $270^\circ$ when the difference due to opposite line directions is neglected. The calculated densities of these dislocations and comparison with the MS results are shown in Fig.~\ref{DislocationDensity_xtoz}(a). It can be seen
that as $\phi$ varies from
from $0^\circ$ when the boundary is the pure twist boundary discussed in Sec.~3.2 to $90^\circ$ when the boundary is a tilt boundary,
 the density of $\mathbf b^{(1)}$-dislocations decreases, while both densities of  $\mathbf b^{(2)}$ and $\mathbf b^{(3)}$-dislocations increase first and then decrease. The predictions of the continuum model on these overall trends of dislocation densities agree with the MS results. Quantitative comparison between the results of the two models shows that the agreement is excellent before $\phi=30^\circ$ and near $\phi=90^\circ$, whereas noticeable deviations exist   near $\phi=60^\circ$.
 These deviations are mainly due to the neglect of the small contribution of interaction energy of crossing dislocations \cite{vitek1987} in the continuum energy formulation, see the dislocation structures to be discussed below and a preliminary modification discussed in Sec.~6.
However, as shown in Fig.~\ref{DislocationDensity_xtoz}(b)  to be discussed below,
the continuum model is still able to give excellent prediction on the grain boundary energy, which is the main purpose of the continuum model.

\begin{figure}[htbp]
\centering
    \includegraphics[width=0.6\linewidth]{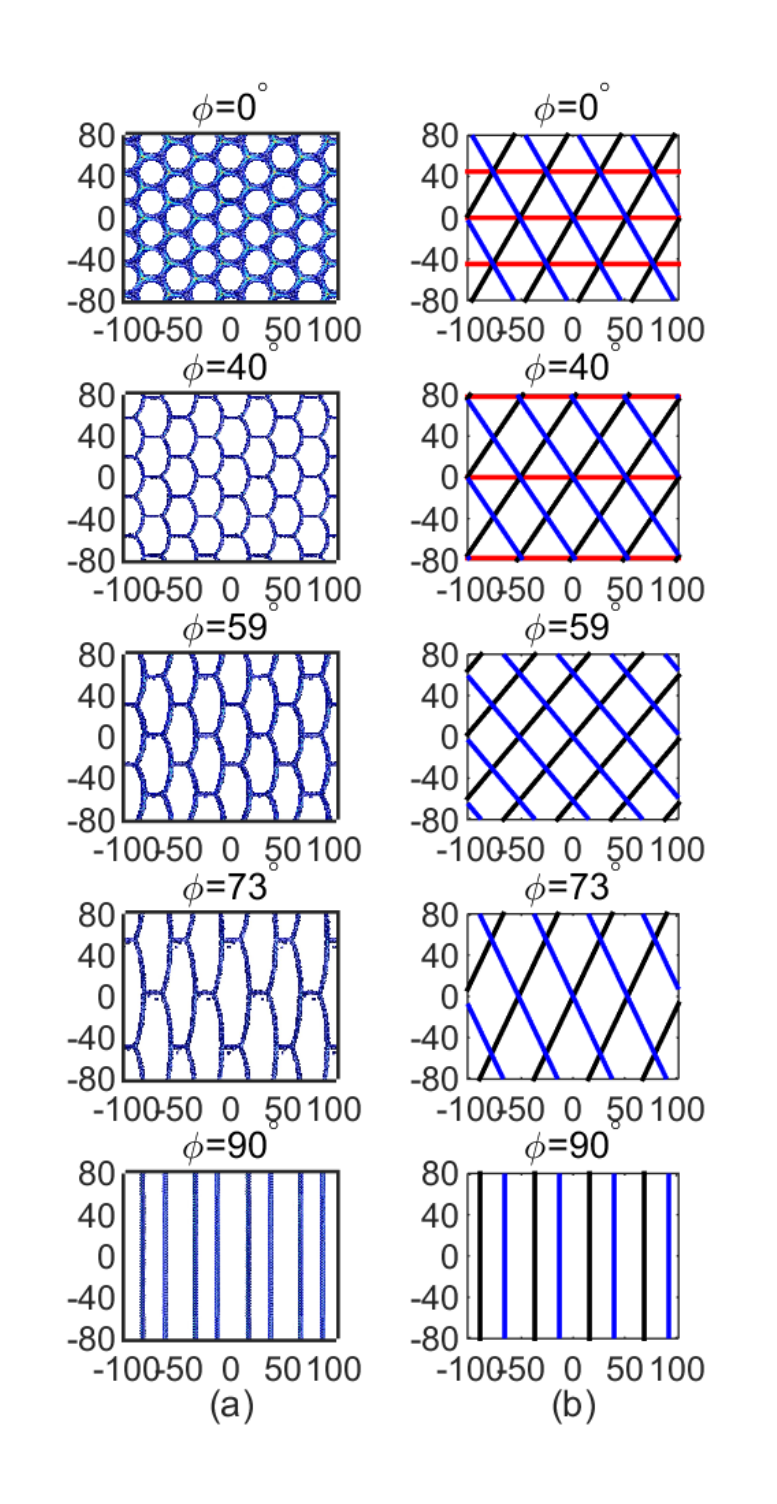}
    \caption{Dislocation structure of the $[111]$ grain boundaries parallel to the $[\bar{1}10]$ direction indexed by the inclination angle $\phi$, computed by (a) MS simulations and (b) the continuum model.  The horizontal  direction is the $[\bar{1}10]$ direction (the $x$ axis in the continuum model).  The misorientation angle is $\theta=1.95^\circ$. The dislocation structures of the grain boundaries with other values of $\phi$ are similar due to the symmetries  discussed in the text.}
    \label{eta_y}
\end{figure}

Fig.~\ref{DislocationDensity_xtoz}(b) shows
 the calculated energy density  of these grain  boundaries  and comparison with the  MS simulation results. We also use the expression of the parameter $r_g$ in Eq.~\eqref{eqn:rg111} fitted in Sec.~4.1 for the continuum energy. Excellent agreement can be seen  between the grain boundary energies computed using these two approaches. The agreement also suggests that the expression of $r_g$ in Eq.~\eqref{eqn:rg111} can be applied more generally to other grain boundaries, which will be further confirmed in the next subsection.
 It can be seen from Fig.~\ref{DislocationDensity_xtoz}(b) that as the inclination angle $\phi$ varies from $0^\circ$ corresponding to the pure twist boundary to $90^\circ$ corresponding to the pure tilt boundary following this path,
the grain boundary energy increases first and then gradually decreases, ending up with a higher energy of the pure tilt boundary than that of the pure twist boundary. The behavior of this grain boundary energy curve is similar to that in Fig.~\ref{DislocationDensity_xtoy}(b) for those grain boundaries along the path (I).

Fig.~\ref{eta_y} shows the obtained dislocation structures on these grain boundaries using the continuum model (Fig.~\ref{eta_y}(b)) and MS simulations (Fig.~\ref{eta_y}(a)). As $\phi$ varies from $0^\circ$ to $90^\circ$, in addition to the variation of dislocation densities shown in Fig.~\ref{DislocationDensity_xtoz}(a), the character of $\mathbf b^{(1)}$-dislocations remains pure screw, and the $\mathbf b^{(2)}$-dislocations and $\mathbf b^{(3)}$-dislocations have more and more edge character from the pure screw character in the twist boundary at $\phi=0^\circ$.
 These agree with the MS results in Fig.~\ref{eta_y}(a).
  For the case of pure tilt boundary with $\phi=90^\circ$ where the constituent dislocations are straight, both the dislocation structure and the dislocation densities computed by the continuum model are in excellent agreement with the MS results, see the last row of images in Fig.~\ref{eta_y} and the dislocation densities in Fig.~\ref{DislocationDensity_xtoz}(a).
 The small deviations in dislocation densities near $\phi=60^\circ$ between the results of the continuum model and MS simulation shown in Fig.~\ref{DislocationDensity_xtoz}(a) can be understood in terms of the dislocation structure shown in Fig.~\ref{eta_y}.  In fact, approximately at  $\phi=54^\circ$,   the $\mathbf b^{(1)}$-dislocations  start to disappear  by the continuum model, see Fig.~\ref{eta_y}(b); whereas in the MS results, dislocation segments with Burgers vector $\mathbf b^{(1)}$ are still present and they only disappear when $\phi=90^\circ$  (the tilt boundary),  see  Fig.~\ref{eta_y}(a). As indicated above, this deviation is mainly caused by the neglect of the  interaction energy of crossing dislocations \cite{vitek1987} in the continuum model. Although this small energy contribution gives noticeable changes in the dislocation densities of these grain boundaries, it does not change the grain boundary energy too much, as can be seen from the comparison in Fig.~\ref{DislocationDensity_xtoz}(b) that the energies computed by the continuum model and the MS simulation are very close to each other. An example will be presented in Sec.~6 showing that including this interaction energy does improve the results of the continuum model.
 Another interesting atomistic feature observed from the MS results is that
 the dislocation segments in the networks on these grain boundaries spread within their own slip planes that are not parallel to the grain boundary planes, leading to jogs on these dislocation segments whose energy is also neglected in the continuum model.

\subsection{$<111>$ grain boundaries parallel to $[111]$ direction}

In this subsection, we consider the $[111]$ grain boundaries parallel to the $[111]$ direction. These grain boundaries are indexed by the inclination angle $\beta$ between the grain boundary normal and the $[\bar{1}\bar{1}2]$ direction, see path (III) in Fig.~\ref{structure21}. These grain boundaries are tilt boundaries. When $\beta=0^\circ$, the grain boundary is the $(\bar{1}\bar{1}2)[111]$ tilt boundary at the end of the path (II) in the previous subsection. When $\beta=90^\circ$, it is  the $[111]$ tilt boundary studied in Sec.~3.1.

\begin{figure}[htbp]
 \centering
   \includegraphics[width=0.9\linewidth]{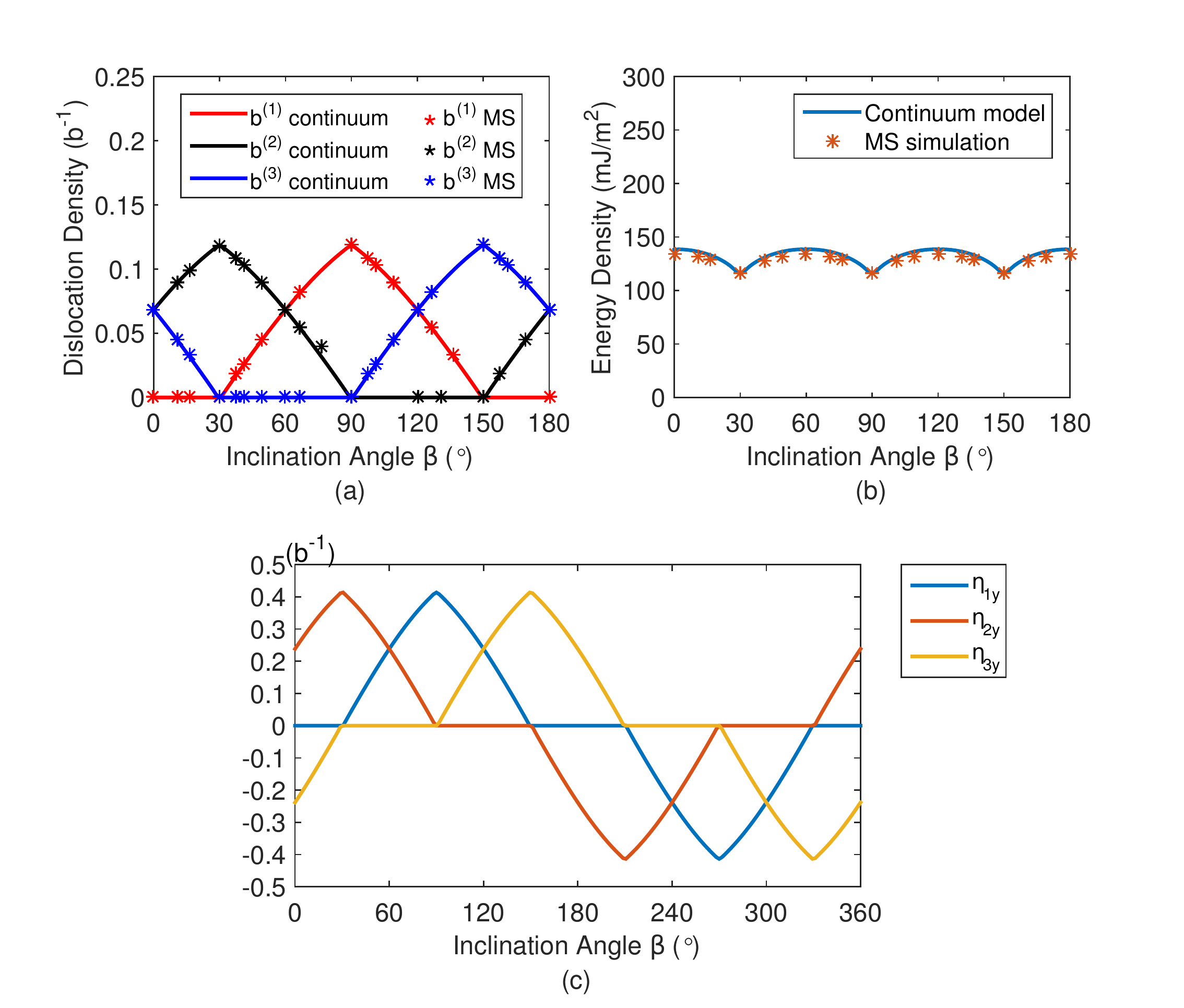}
    \caption{Dislocation densities (a) and grain boundary energy density (b) of the $[111]$ grain boundaries parallel to the $[111]$ direction indexed by the inclination angle $\beta$, computed by the continuum model and MS simulations. The misorientation angle is $\theta=1.95^\circ$. (c) The converged values of  $\eta_{jx}$ and $\eta_{jy}$ in our continuum model for the dislocation structure on these grain boundaries. The $x$ axis in the continuum model is always in the $[111]$ direction. Note that except for $\eta_{jy}$, $j=1,2,3$, the remaining components are $0$.  } \label{DislocationDensity_ytoz}
\end{figure}

The values of  $\eta_{jx}$ and $\eta_{jy}$, $j=1,2,\cdots,6$, on these grain boundaries obtained  using the continuum model are shown in Fig.~\ref{DislocationDensity_ytoz}(c) in terms of the inclination angle $\beta$. The dislocation structure of these grain boundaries also consists of dislocations with Burgers vectors $\mathbf b^{(1)}$, $\mathbf b^{(2)}$,  and $\mathbf b^{(3)}$.
These dislocations are all in the $[111]$ direction (the $x$ axis), indicated by the result $\eta_{jx}=0$ for $j=1,2,3$.
The following symmetry can be seen that about $\beta=90^\circ$ and $180^\circ$, the dislocation distributions are symmetric if we interchange the profiles of $\mathbf b^{(2)}$-dislocations and $\mathbf b^{(3)}$-dislocations (neglecting the difference due to opposite line directions).  The calculated densities of these dislocations and comparison with the MS results are shown in Fig.~\ref{DislocationDensity_ytoz}(a). Again our continuum model is able to give excellent approximations to the densities of dislocations on these grain boundaries. The results of both models show that there is a phase shift of $60^\circ$ from the distribution of $\mathbf b^{(1)}$-dislocations to that of $\mathbf b^{(2)}$-dislocations,  and from that of  $\mathbf b^{(2)}$-dislocations to that of  $\mathbf b^{(3)}$-dislocations. The period of these dislocation densities is  $\beta=180^\circ$.

\begin{figure}[htbp]
\centering
    \includegraphics[width=.65\linewidth]{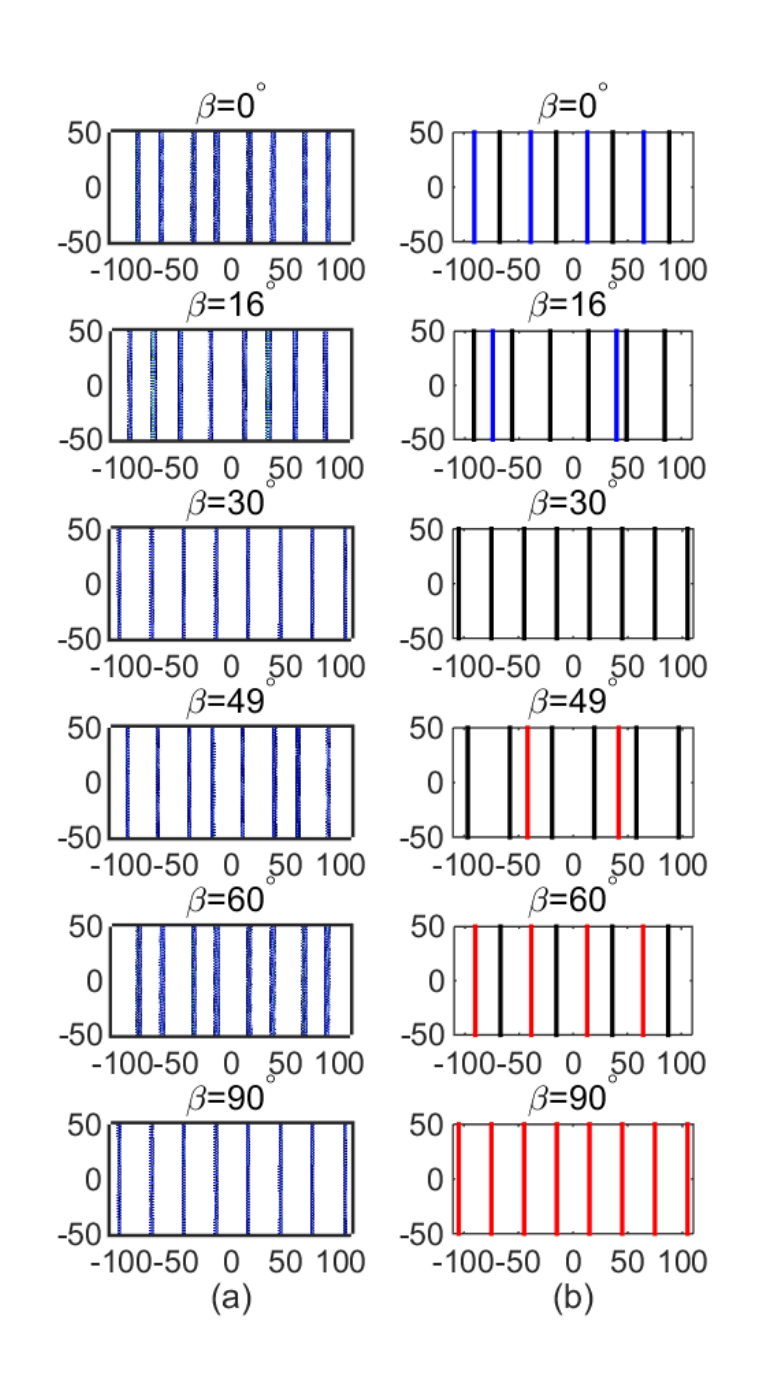}
     \caption{Dislocation structure of the $[111]$ grain boundaries parallel to the $[111]$ direction indexed by the inclination angle $\beta$ as shown in Fig.~\ref{structure21}, computed by (a) MS simulations and (b) the continuum model.  The vertical  direction is the $[111]$ direction (the $x$ axis in the continuum model).  The misorientation angle is $\theta=1.95^\circ$. The dislocation structures of the grain boundaries with other values of $\phi$ are similar due to the symmetries  discussed in the text.}
    \label{eta_x}
\end{figure}

Fig.~\ref{eta_x} shows the obtained dislocation structures on these grain boundaries using the continuum model (Fig.~\ref{eta_x}(b)) and MS simulations (Fig.~\ref{eta_x}(a)). We can see that the continuum model is able to give accurate dislocation structures compared with the MS results
for these networks of straight dislocations. As discussed above, dislocations on these boundaries are always parallel to the rotation axis $[111]$. Combined with the dislocation densities shown in Fig.~\ref{DislocationDensity_ytoz}(a), it can be seen  that when $\beta=30^\circ$, $90^\circ$, and $150^\circ$, the grain boundary is a symmetric tilt boundary that consists only of $\mathbf b^{(2)}$-dislocations, $\mathbf b^{(1)}$-dislocations,  and $\mathbf b^{(3)}$-dislocations, respectively.
In the interval $\beta\in[30^\circ,90^\circ]$, the density of $\mathbf b^{(1)}$-dislocations increases almost linearly, the density of $\mathbf b^{(2)}$-dislocations  decreases almost linearly, and the density of $\mathbf b^{(3)}$-dislocations remains $0$. Behaviors of dislocation densities on other $60^\circ$-intervals are the same except that the roles played by dislocations with different Burgers vectors rotate among themselves.

Fig.~\ref{DislocationDensity_ytoz}(b) shows
 the calculated energy density  of these grain  boundaries and comparison with MS simulation results.
 We still use the expression of the parameter $r_g$ in Eq.~\eqref{eqn:rg111} in the continuum model, where the azimuthal angle $\phi=90^\circ$  and $r_g=0.85b$ for all values of $\beta$. Excellent agreement can be seen  between the grain boundary energies computed using these two approaches. The agreement confirms again that the expression of $r_g$ in Eq.~\eqref{eqn:rg111} can be applied generally to all $<111>$ grain boundaries.
It can be seen from Fig.~\ref{DislocationDensity_ytoz}(b) that the energy density of these tilt boundaries have a period of $60^\circ$ in the inclination angle $\beta$, which agrees with the behaviors of dislocation densities and structure discussed above and the crystallography shown in Fig.~\ref{structure21}. The energy has minima at $\beta=30^\circ$, $90^\circ$, and $150^\circ$ corresponding to the $<111>(110)$ symmetric tilt boundaries that consist only of $\mathbf b^{(2)}$-dislocations, $\mathbf b^{(1)}$-dislocations,  and $\mathbf b^{(3)}$-dislocations, respectively.

\subsection{$<111>$ low angle grain boundaries with any boundary plane orientation}

\begin{figure}[htbp]
 \centering
    \includegraphics[width=0.9\linewidth]{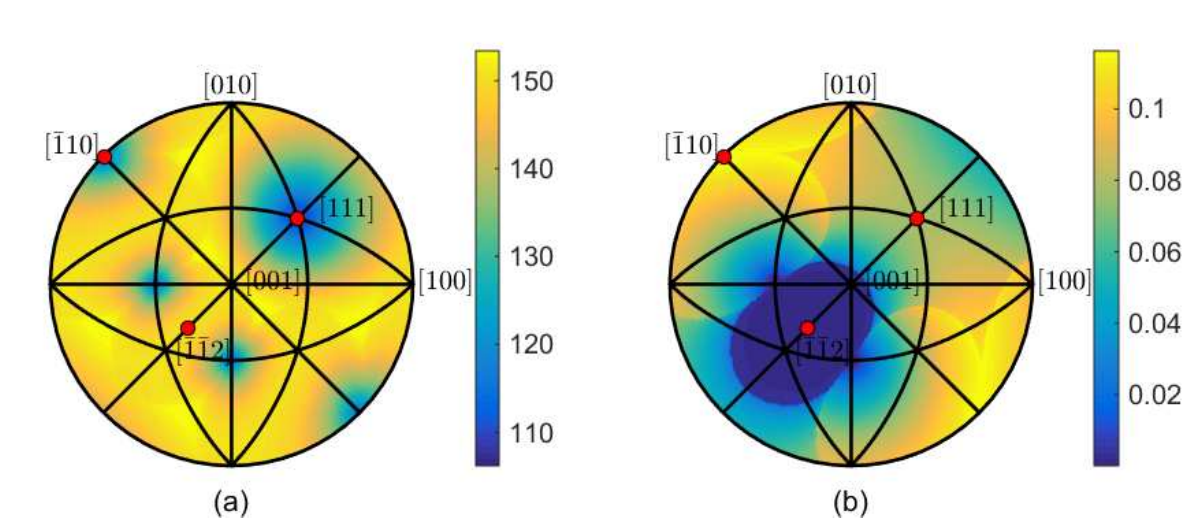}
    \caption{Grain boundary energy (a) and density of $\mathbf b^{(1)}$-dislocations (b) of $[111]$ low angle grain boundaries varies with boundary plane orientation in stereographic projection. Densities of $\mathbf b^{(2)}$ and $\mathbf b^{(3)}$-dislocations can be obtained by rotating the density of $\mathbf b^{(1)}$-dislocations in (b) by $60^\circ$ and $120^\circ$ around the $[111]$ direction. The misorientation angle of these grain boundaries is  $\theta=1.95^\circ$. The energy unit is $mJ/m^2$. The $(111)$, $(\bar{1}10)$, and $(\bar{1}\bar{1}2)$ grain boundaries are the starting and ending states of the paths of boundaries studied in the previous subsections. }\label{fig:energy111}
\end{figure}

In this subsection, using our continuum model  with parameter $r_g$ given by Eq.~\eqref{eqn:rg111}, we calculate the energy of all low angle grain boundaries as a function of the boundary plane orientation, with fixed rotation axis $\mathbf a=[111]$ and misorientation angle $\theta=1.95^\circ$.
The energy densities of these low angle grain boundaries are shown in Fig.~\ref{fig:energy111}(a) in stereographic projection. The minimum energy states are  the $(111)$ twist boundaries. Local energy minima are achieved at the $(\bar{1}10)$, $(\bar{1}01)$, and $(0\bar{1}1)$  symmetric tilt boundaries. The grain boundaries with the maximum energy  have the orientations $(\bar{1}13)$, $(1\bar{1}3)$, $(\bar{1}31)$, $(\bar{1}\bar{3}1)$, $(3\bar{1}1)$, and $(\bar{3}\bar{1}1)$.

Densities of $\mathbf b^{(1)}$-dislocations on
 these $[111]$ low angle grain boundaries are shown in Fig.~\ref{fig:energy111}(b).
 Densities of $\mathbf b^{(2)}$ and $\mathbf b^{(3)}$-dislocations can be obtained by rotating the density of $\mathbf b^{(1)}$-dislocations  by $60^\circ$ and $120^\circ$ around the $[111]$ direction (see Fig.~\ref{structure21}).
 Dislocations with the remaining Burgers vectors do not appear on these boundaries.
  The maximum densities of these dislocations appear in the $(\bar{1}10)$, $(\bar{1}01)$, and $(0\bar{1}1)$  symmetric tilt boundaries, where  $\mathbf b^{(1)}$, $\mathbf b^{(2)}$, and $\mathbf b^{(3)}$ are the solo dislocations in the networks, respectively. Another observation is that on those  $(\bar{1}13)$, $(1\bar{1}3)$, $(\bar{1}31)$, $(\bar{1}\bar{3}1)$, $(3\bar{1}1)$, and $(\bar{3}\bar{1}1)$ boundaries that have the highest energy, the densities of dislocations of all the three Burgers vectors are nonzero.

\section{Low angle grain boundaries as rotation axis varies}

In this section, we calculate the grain boundary  energy using our continuum model with varying rotation axis $\mathbf a$ and fixed grain boundary plane orientation $\mathbf n=(111)$ and misorientation angle  $\theta=1.95^\circ$.
The rotation axis $\mathbf a$ varies following the path (I) in Fig.~\ref{structure21}.

\begin{figure}[htbp]
 \centering
   \includegraphics[width=0.9\linewidth]{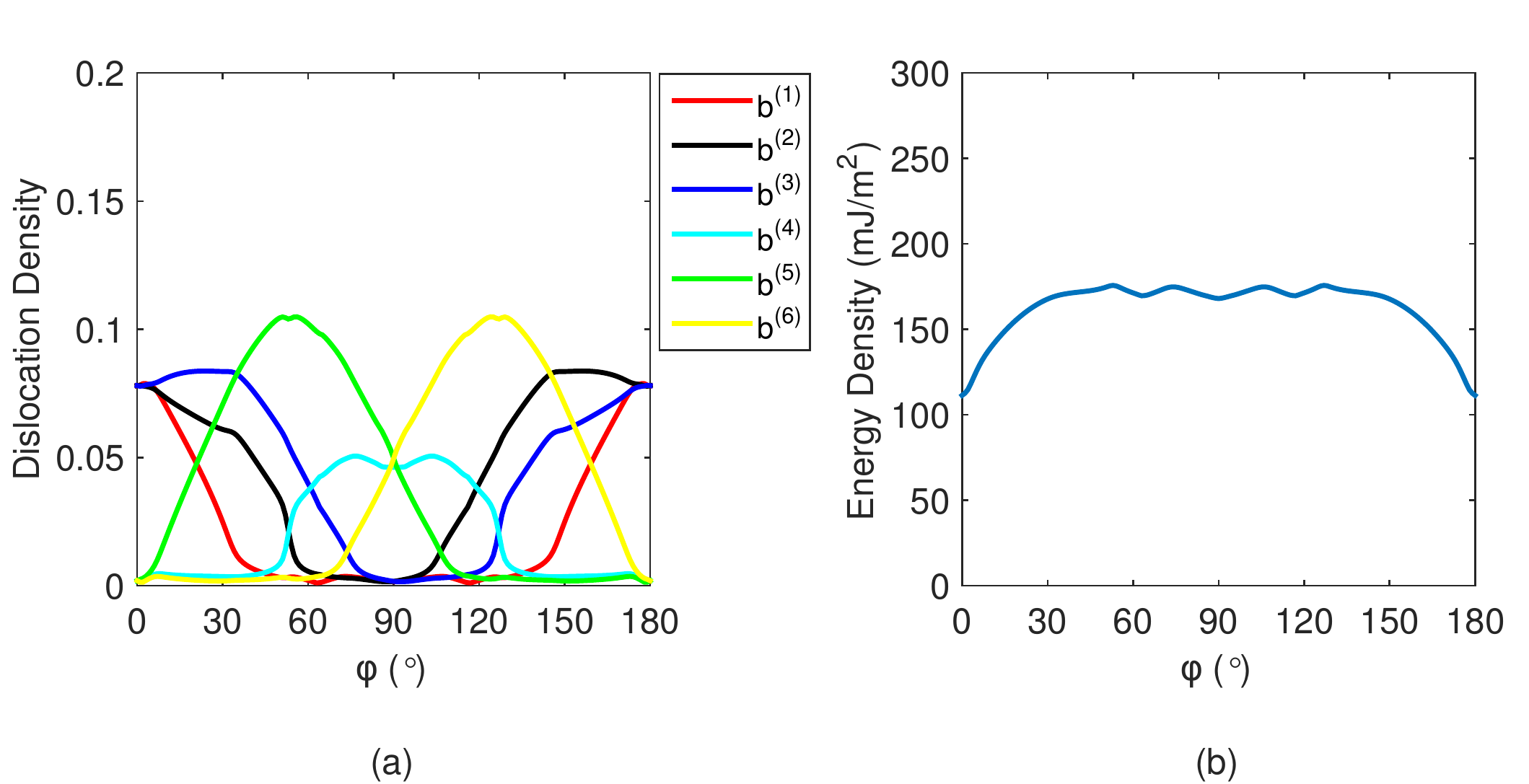}
    \caption{Dislocation densities (a) and grain boundary energy density (b) of the grain boundaries
    with rotation axis $\mathbf a$ varying along the path (I) in Fig.~\ref{structure21} computed by the continuum model. The grain boundary plane orientation is fixed to be $\mathbf n=(111)$  and the misorientation angle  is $\theta=1.95^\circ$. This grain boundaries are indexed by the azimuthal angle $\phi$ between the rotation axis $\mathbf a$ and the $[111]$ direction.} \label{RA_density_energy}
\end{figure}

Fig.~\ref{RA_density_energy} shows the dislocation densities and grain boundary energy density of these grain boundaries, indexed by the azimuthal angle $\phi$ between the rotation axis $\mathbf a$ and the $[111]$ direction. A direct observation from the dislocation densities shown in Fig.~\ref{RA_density_energy}(a) is that dislocations with all the six Burgers vectors appear in the dislocation networks on the grain boundaries as the rotation axis $\mathbf a$ varies along this path. More interestingly,  as the rotation axis $\mathbf a$ varies from $\phi=0^\circ$ to $90^\circ$, the $\mathbf b^{(1)}$-dislocations is gradually replaced by the  $\mathbf b^{(5)}$-dislocations, then the $\mathbf b^{(2)}$-dislocations are  replaced by the  $\mathbf b^{(4)}$-dislocations, and finally the $\mathbf b^{(3)}$-dislocations are replaced by the  $\mathbf b^{(6)}$-dislocations. The combination of the Burgers vectors of the constituent dislocations has the pattern
$\{\mathbf b^{(1)}, \mathbf b^{(2)}, \mathbf b^{(3)}, \mathbf b^{(5)}\}$,
$\{\mathbf b^{(2)}, \mathbf b^{(3)}, \mathbf b^{(5)}\}$,
$\{\mathbf b^{(2)}, \mathbf b^{(3)}, \mathbf b^{(4)}, \mathbf b^{(5)}\}$,
$\{\mathbf b^{(3)}, \mathbf b^{(4)}, \mathbf b^{(5)}\}$,
$\{\mathbf b^{(3)}, \mathbf b^{(4)}, \mathbf b^{(5)}, \mathbf b^{(6)}\}$, and
$\{\mathbf b^{(4)}, \mathbf b^{(5)}, \mathbf b^{(6)}\}$.
 Recall that we start from the pure twist boundary consisting of dislocations with Burgers vectors $\{\mathbf b^{(1)}, \mathbf b^{(2)}, \mathbf b^{(3)}\}$ at $\phi=0^\circ$. When $\phi=90^\circ$, the grain boundary is a tilt boundary consisting of dislocations $\{\mathbf b^{(4)}, \mathbf b^{(5)}, \mathbf b^{(6)}\}$, and these dislocations have the same density and are parallel straight lines.
The dislocation densities have symmetries beyond $\phi=90^\circ$ if the roles of some dislocations are interchanged. The energy of these grain boundaries is shown in Fig.~\ref{RA_density_energy}(b), which has a minimum value for the twist boundary at  $\phi=0^\circ$ and local minimum value for the tilt boundaries at $\phi=90^\circ$.

\section{Conclusions and discussion}

In this paper,  we present a continuum model to compute the  energy of low angle grain boundaries for any given values of the five DOFs. In our continuum model, we minimize the grain boundary energy associated with the dislocation structure subject to the constraint of Frank's formula for dislocations with all possible Burgers vectors. This constrained minimization problem is solved by the penalty method, which turns the problem into an unconstrained minimization problem.
The grain boundary dislocation structure is approximated by a network of straight dislocations that predicts the energy and dislocation densities of the grain boundaries, and incorporates the anisotropic nature in the grain boundary energy.
  This continuum model can be considered a generalization of the classical Read-Shockley model to the cases in which the Frank's formula is satisfied by multiple  dislocation structures.

We use our continuum model to systematically study the dislocation densities and grain boundary energy of low angle grain boundaries in fcc Al with rotation axis in the $[111]$ direction and any boundary plane orientation. The results are examined by MS simulation results. Comparisons with the MS results show that our continuum model is able to provide good predictions of the densities of the constituent dislocations and the anisotropic energy density of low angle grain boundaries. An expression of the core parameter $r_g$ in the energy formula is obtained which depends only on the azimuthal angle of the grain boundary normal with respect to the rotation axis. We also calculate the grain boundary  energy using our continuum model with varying rotation axis $\mathbf a$ and fixed grain boundary plane orientation and misorientation angle.

\begin{figure}[htbp]
 \centering
   \includegraphics[width=0.9\linewidth]{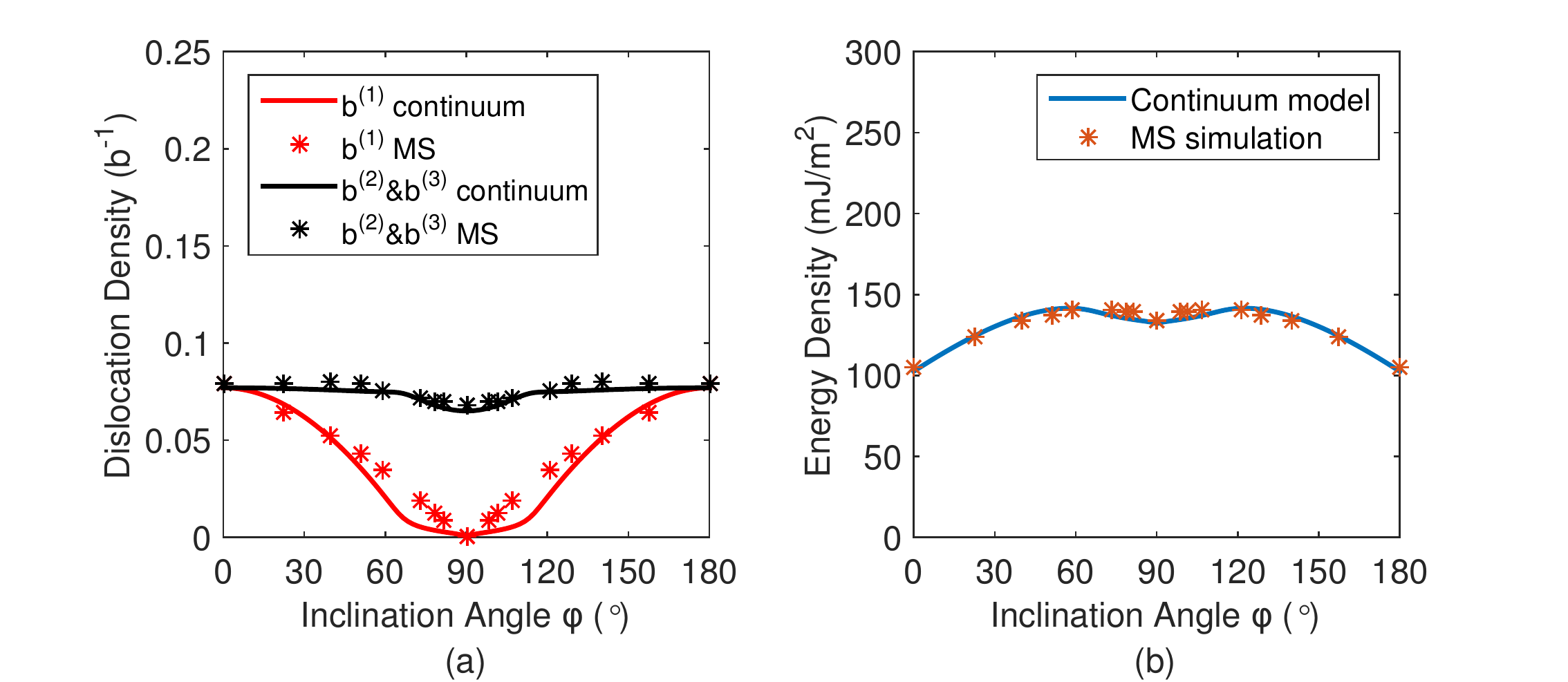}
    \caption{Dislocation densities (a) and grain boundary energy density (b) for those grain boundaries considered in Fig.~\ref{DislocationDensity_xtoz}, computed by the continuum model with the modified grain boundary energy formula in Eq.~\eqref{eqn:modify} and MS simulations.   } \label{density_energy2_discussion}
\end{figure}

We have seen in Fig.~\ref{DislocationDensity_xtoz} that for the grain boundaries along the path (II) in Fig.~\ref{structure21}, there are some small deviations in dislocation densities obtained by our continuum model compared with the MS results. We have attributed these deviations mainly to the neglect of interaction energy of the crossing dislocations in the networks on these grain boundaries. Here we perform some preliminary examinations to see if such deviations can be reduced by incorporating this interaction energy. Suggested by the work based on the interaction energy of dislocation networks on twist boundaries \cite{vitek1987}, we add a gradient-squared term in the grain boundary energy in Eq.~\eqref{eqn:gb_density} to account for this energy contribution phenomenologically. The modified grain boundary energy takes the form
\begin{equation}
\gamma_{\rm gb}^{\rm mod}={\displaystyle \sum_{j=1}^J  \left\{ \frac{\mu(b^{(j)})^2}{4\pi(1-\nu)}\!\left[1-\nu\frac{(\nabla \eta_j\! \times \!\mathbf{n} \!\cdot\! \mathbf{b}^{(j)})^2}{(b^{(j)})^2 {\|\nabla \eta_j\|}^2}\right]\!\|\nabla \eta_j\| \log\! \frac{1}{r_g\sqrt{\|\nabla \eta_j\|^2+\epsilon}}+\lambda \|\nabla \eta_j\|^2 \right\}},\label{eqn:modify}
\end{equation}
where $\lambda$ is some positive parameter.
Simulation results using this modified energy are shown in Fig.~\ref{density_energy2_discussion} with
$\lambda=5\mu b^3/2(1-\nu)$. We can see that the dislocation densities calculated by the continuum model have been improved significantly, and the calculated grain boundary energy becomes more accurate with the minimum at $\phi=90^\circ$ being clearly seen. These preliminary results confirm the conclusion on the origin of the small deviations in dislocation densities in Fig.~\ref{DislocationDensity_xtoz}. It is expected that the exact energy formulation that accounts for such dislocation interaction may further improve the results of the continuum model, especially the density of  $\mathbf{b}^{(1)}$-dislocations near $\phi=90^\circ$ in Fig.~\ref{density_energy2_discussion}(a). Moreover, as reported in Sec.~4.2, the energy of jogs on the dislocation segments  observed in the atomistic results in Fig.~\ref{eta_y}(a) is neglected in the continuum model. Incorporation of this energy of jogs may also improve the continuum model.
The formulas of these improvements are being derived and the results will be reported elsewhere.

  The formulations of the constrained energy minimization problem in Eqs.~\eqref{eqn:gb_energy}-\eqref{eqn:frank} and the unconstrained-problem approximation in Eq.~\eqref{penalty} for determining the energy of low angle grain boundaries also apply when the constituent dislocations are curved and/or the grain boundaries are curved. When the grain boundary is curved, $\nabla \eta$ in Eq.~\eqref{eqn:gb_density} will be replaced by the surface gradient $\nabla_S\eta = [\nabla-\mathbf{n}(\mathbf{n}\cdot\nabla)]\eta$, where $\mathbf{n}$ is the local unit normal vector of the grain boundary \cite{Zhu2014175}. The dislocation structure obtained using our continuum model can also be used as initial conditions in the  discrete dislocation dynamics simulations for the dislocation structure of low angle grain boundaries that have higher resolution than the continuum model \cite{Lim20095013,Lim20121395,Quek2011,Winther2013,Winther2014}.
   The continuum model presented in this paper is based on the dislocation model of low angle grain boundaries. At higher misorientation angles, the core regions of the constituent dislocations of the grain boundaries heavily overlap. These high angle grain boundaries can be considered as some known structures plus arrays of secondary grain boundary dislocations \cite{Sutton1995}, and generation of our continuum model can be made accordingly.  In this paper, the continuum model is applied to fcc crystals (Al), it can also be applied to grain boundary energy and structure in other crystals such as hcp \cite{Wang1-2012,Wang2-2012,WangJianSciRep2016} where molecular dynamics results are available for some special grain boundaries. Grain boundary motion and grain growth can also be studied with the obtained anisotropic grain boundary energy \cite{Sutton1995}.  With  modifications, our model can apply to  dislocation structures and energy of heterogeneous interfaces \cite{Quek2011,Hirth2013,WangJian2013,Demkowicz2013},
    including those in nanolayered composites in which these interfaces have shown to  play crucial roles on the mechanical and plastic behaviors of these materials \cite{WangJianSciRep2013,Salehinia2014,Salehinia2015,Shao2015}. These generalizations and applications will be explored in the future work.

\section*{Acknowledgments}
The authors would like to thank Professor David J. Srolovitz of University of Pennsylvania for helpful discussions. This work was partially supported by the Hong Kong Research Grants Council General Research Fund 606313.

\end{document}